\documentclass[12pt]{article}

\usepackage{amssymb}
\usepackage{amsmath}
\usepackage{amscd}
\usepackage{latexsym}

\usepackage{graphicx}
\usepackage{epstopdf}
\usepackage{psfrag}                  
\usepackage{url}                        
\usepackage{times}                    
\usepackage{makeidx}               
\usepackage{multicol}                
\usepackage[bottom]{footmisc} 
\usepackage{fancyhdr}
\RequirePackage{subfigure}
\usepackage[square, comma, sort&compress]{natbib}

\newcommand{\be}{\begin{equation}}
\newcommand{\ee}{\end{equation}}

\newcommand{\dlt}{\delta}

\newcommand{\ra}{\rightarrow}
\newcommand{\sgm}{\sigma}

\newcommand{\gm}{\gamma}

\newcommand{\lbd}{\lambda}

\begin{document}

\begin{center}

{\Large {\bf Punctuated Evolution due to Delayed
Carrying Capacity} \\ [5mm]

V.I. Yukalov$^{1,2}$, E.P. Yukalova$^{1,3}$, and
D. Sornette$^{1,4}$ } \\ [3mm]

{\it $^1$Department of Management, Technology and Economics, \\
ETH Z\"urich, Z\"urich CH-8032, Switzerland, \\ [2mm]

$^2$Bogolubov Laboratory of Theoretical Physics, \\
Joint Institute for Nuclear Research, Dubna 141980, Russia, \\ [2mm]

$^3$Laboratory of Information Technologies, \\
Joint Institute for Nuclear Research, Dubna 141980, Russia, \\ [2mm]

$^4$ Swiss Finance Institute, \\
c/o University of Geneva, 40 blvd. Du Pont d'Arve,
CH 1211 Geneva 4, Switzerland}

\end{center}

\vskip 3cm

\begin{abstract}

A new delay equation is introduced to describe the
punctuated evolution of complex nonlinear systems. A detailed
analytical and numerical investigation provides the classification
of all possible types of solutions for the dynamics of a population
in the four main regimes dominated respectively by: (i) gain and
competition, (ii) gain and cooperation, (iii) loss and competition
and (iv) loss and cooperation. Our delay equation may
exhibit bistability in some parameter range, as well as a rich set
of regimes, including monotonic decay to zero, smooth exponential
growth, punctuated unlimited growth, punctuated growth or alternation
to a stationary level, oscillatory approach to a stationary level,
sustainable oscillations, finite-time singularities as well
as finite-time death.

\end{abstract}

\vskip 2cm

{\bf Pacs}: 02.30.Ks, 87.23.Ce, 87.23.Ge, 87.23.Kg, 89.65.Gh

\vskip 2cm

{\bf Keywords}: Delay equations; Population dynamics; Punctuated
evolution

\newpage


\section{Introduction}

Most natural and social systems evolve according to multistep
processes. We refer to this kind of dynamics as {\it punctuated
evolution}, because it describes the behavior of nonequilibrium
systems that evolve in time, not according to a smooth or gradual
fashion, but by going through periods of stagnation interrupted
by fast changes. These include the growth of urban population
\cite{13,14}, the increase of life complexity and the development of
technology of human civilizations  \cite{15}, and, more prosaically,
the natural growth of human bodies \cite{16}.

According to the theory of punctuated equilibrium
\cite{17,18,19,20,21}, the
evolution of the majority of sexually reproducing biological
species on Earth also goes through a series of sequential
growth-stagnation stages. For most of their geological history,
species experience little morphological change. However, when
phenotypic variation does occur, it is temporally localized in
rare, rapid events of branching speciation, called cladogenesis;
these rapid events originate from genetic revolutions by
allopatric and peripatric speciations \cite{22,23,24}. The resulting
punctuated-equilibrium concept of the evolution of biological
species is well documented from paleontological fossil records
\cite{17,18,19,20,21,25,26}. It does not contradict the Darwin's theory of
evolution \cite{27}, but rather emphasizes that evolution processes
do not unfold continuously and regularly. The change rates vary
with time, being almost zero for extended geological periods,
and strongly increasing for short time intervals. This supports
Darwin's remark \cite{27} that ``each form remains for long periods
unaltered, and then again undergoes modification.'' Here ``long''
and ``short" are to be understood in terms of geological time
scale, with ``long" meaning hundreds of millions of years and
``short" corresponding to thousands or hundreds of thousands
of years.

The development of human societies provides many other examples
of punctuated evolution. For instance, governmental policies,
as a result of bounded rationality of decision makers \cite{28},
evolve incrementally \cite{29}. The growth of organizations, of
firms, and of scientific fields also demonstrates nonuniform
developments, in which relatively long periods of stasis are
followed by intense periods of radical changes \cite{30,31,32}.
During the training life of an athlete, sport achievements rise
also in a stepwise fashion \cite{33}.

Despite these ubiquitous empirical examples of punctuated
evolution occurring in the  development of many evolving
systems, to our knowledge, there exists no mathematical
model describing this kind of evolution. It is the aim of
the present paper to propose such a mathematical model, which
is very simple in its structure and its conceptual foundation.
Nevertheless, it is surprisingly rich in the variety of regimes
that it describes, depending on the system parameters. In
addition to the process of punctuated increase, it demonstrates
punctuated decay, punctuated up-down motion, effects of mass
extinction, and finite-time catastrophes.

The paper is organized as follows. Section 2 presents the
derivation of the novel logistic delay equation that we study
in the rest of the paper. Section 3 describes the methodology
used to study the logistic delay equation, both analytically
and numerically. The four following sections 4-7 present the
classification of all possible types of solutions for the
dynamics of a population obeying our logistic delay equations,
analyzing successively the four possible situations dominated
respectively by: (i) gain and competition, (ii) gain and
cooperation, (iii) loss and competition, and (iv) loss and
cooperation. Section 8 concludes by providing figures, 
which summarize all possible regimes.


\section{Model formulation}

\subsection{Derivation of the general model}

The logistic equation, advanced by Verhulst \cite{1}, has been 
the workhorse model for describing the evolution of various 
social, biological, and economic systems:
\begin{equation}
\frac{dN(t)}{dt} = r N(t) \left[ 1 \; - \; \frac{N(t)}{K}
\right] \; .
\label{rhgejv}
\end{equation}
Here $N(t)$ is a measure characterizing the system development,
e.g., the population size, the penetration of new commercial
products or the available quantity of assets. The coefficient
$r$ is a reproduction rate and $K$ is the carrying capacity.
The expression $r (1 - N/K)$ is interpreted as an effective
reproduction rate which, in expression (\ref{rhgejv}), adjusts
instantaneously to $N(t)$. It is possible to assume that this
effective reproduction rate lags with a delay time $\tau$,
leading to the suggestion by Hutchinson \cite{2}  to consider
the equation
\begin{equation}
\label{thtqe;q}
\frac{dN(t)}{dt} = r N(t) \left[ 1 \; - \;
\frac{N(t-\tau)}{K} \right] \; ,
\end{equation}
termed the delayed logistic equation. Many other variants
of the logistic equation have been proposed \cite{3,4,5,6,7,8,9},
uniform or nonuniform, with continuous or discrete time, and
with one or several delays. An extensive literature on such
equations can be found in the books \cite{10,11,12}.

All known variants of the logistic equation describe
either a single-step evolution, called the $S$-curve, or
an oscillatory behavior around a constant level. However,
as summarized in the introduction, the development of many
complex systems consists not just of a single step, where
a period of fast growth is followed by a lasting period of
stagnation or saturation. Instead, many systems exhibit a
succession of $S$-curves, or multistep growth phases, one
fast growth regime followed by a consolidation, which is
itself followed by another fast growth regime, and so on.
This multistep precess can be likened to a staircase with
approximately planar plateaus interrupted by rising steps.

Motivated by the ubiquity of the multistep punctuated
evolution dynamics on the one hand and the simplicity of
the logistic equation on the other hand, we now propose
what, we think, is the simplest generalization of the
logistic equation that allows us to capture the previously
described phenomenology and much more.

Our starting point is to take into account the main two
causes of development, (i) the evolution of separate
individuals composing the system and (ii) their mutual
collective interactions, leading to the consideration of
two terms contributing to the rate of change of $N(t)$:
\be
\label{eq1}
\frac{dN(t)}{dt} = \gm N(t) \; - \;
\frac{CN^2(t)}{K(t)}  \; .
\ee
The first term $\gm N(t)$ embodies the individual balance
between birth and death, or gain and loss (depending on
whether a population size or economic characteristics are
considered), i.e., the growth rate can be written
\be
\label{eq2}
\gm = \gm_{birth} - \gm_{death} = \gm_{gain} -
\gm_{loss} \; .
\ee
The second term $CN^2(t)/K(t)$ describes collective effects,
 with the coefficient $C$ defining the balance between
competition and cooperation,
\be
\label{eq3}
C = C_{comp} - C_{coop} \; .
\ee
The denominator in the second term of Eq. (\ref{eq1}) can
be interpreted as a generalized carrying capacity.

The principal difference between Eq. (\ref{eq1}) and the
logistic equation is the assumption that the carrying
capacity is a function of time. We assume that the carrying
capacity is not a simple constant describing the available
resources, but that these resources are subjected to the
change due to the activity of the system individuals, who
can either increase the carrying capacity by creative work
or decrease it by destructive actions. Given the co-existence
of both creative and destructive processes impacting the
carrying capacity $K(t)$, we formulate it as the sum of two
different contributions:
\be
\label{eq4}
K(t) = A + B N(t-\tau) \; .
\ee
The first term $A$ is the pre-existing carrying capacity,
e.g., provided by Nature. In contrast, the second term is
the capacity created (or destroyed) by the system. To fix
ideas, let us illustrate by using this model the evolution
of human population of the planet Earth. Then, the second
term $B N(t-\tau)$ is meant to embody the delayed impact
of past human activities in the present services provided
by the planet. There are many complex feedback loops
controlling how human activities interact with the planet
regeneration processes and it is generally understood that
these feedback loops are not instantaneous but act with
delays. A full description of these phenomena is beyond
the scope of this paper. For our purpose, we encapsulate
the complex delayed processes by a single time lag $\tau$,
which will be one of the key parameters of our model. We
stress that  the delay time $\tau$ is introduced to describe
the impact of past human activity on the present value of
the carrying capacity. This is crucially different from the
description (\ref{thtqe;q}) by Hutchinson \cite{2} and others,
in which $\tau>0$ represents delayed interactions between
individuals. In our model  (\ref{eq1}) with (\ref{eq4}),
the cooperation and competition between individuals are
controlled by instantaneous interactions $N(t)\times N(t)$,
while the present carrying capacity $K(t)$ reflects the
impact of the population in the past at time $t-\tau$. The
lag time $\tau$ is thought of as embodying a typical time
scale for regeneration or decay of the renewable resources
provided by the planet. If positive (respectively, negative),
the parameter $B$ describes a productive (respectively,
destructive) feedback of the population on the carrying
capacity.

Although Eq. (\ref{eq1}) with (\ref{eq4}) is reminiscent
of the logistic equation (\ref{rhgejv}), it is qualitatively
different from it by the existence of the time-dependent
delayed carrying capacity. As we show in the sequel, this
difference turns out to be mathematically extremely important,
leading to a variety of evolution regimes that do not exist
in the logistic equation, neither in the standard version nor
in the delayed one of Hutchinson \cite{2} and others.

In particular, the delayed response of the carrying capacity
to the population dynamics is found to be responsible for the occurrences
of regimes in which growth or decay unfold jerkily in a series
of stagnations interrupted by fast changes. The duration
of these plateaus is controlled by the characteristic delay time
scale $\tau$, which can be arbitrarily long or short depending on the 
domain of application.  We capture
this remarkable phenomenon by using the term
{\it punctuated  evolution} in the title (in
contrast with ``punctuated equilibrium'').

\subsection{Reduced variables and parameters}

The quantity $N(t)$ is always measured in some units
$N_{eff}$. For instance, this could be millions of persons
when population is considered, or billions of currency units
for economic systems, or thousands of tons  of goods for
firm production. Hence, it is reasonable to define the
relative quantity
\be
\label{eq9}
x(t) \equiv \frac{N(t)}{N_{eff} } \; .
\ee
For mathematical analysis, it is convenient to deal with
dimensionless quantities. We thus define the dimensionless
value of the pre-existing resources
\be
\label{eq10}
a \equiv \frac{A}{N_{eff}} \; \left | \frac{\gm}{C} \right |
\ee
and of the production (or destruction) factor
\be
\label{eq11}
b \equiv B \; \left | \frac{\gm}{C} \right | \; .
\ee
The dimensionless carrying capacity is defined as
\be
\label{eq12}
y(t) \equiv \frac{K(t)}{N_{eff}} \; \left | \frac{\gm}{C}
\right | \; .
\ee
Using (\ref{eq10}) and (\ref{eq11}), expression (\ref{eq4})
becomes
\be
\label{eq13}
y(t) = a + b x(t-\tau) \; .
\ee
We also define the following notation for the signs of
$\gamma$ and of $C$:
\be
\label{eq14}
\sgm_1 \equiv \frac{\gm}{|\gm|} \; , \qquad \sgm_2 \equiv
\frac{C}{|C|} \; .
\ee
Using a time measured in unit of $1/\gamma$, and keeping the
same notation of time, Eq. (\ref{eq1})
reduces to the evolution equation for $x=x(t)$,
\be
\label{eq15}
\frac{dx}{dt} = \sgm_1 x \; - \; \sgm_2\; \frac{x^2}{y} \; ,
\ee
in which $y = y(t)$ is given by Eq. (\ref{eq13}).

There are two time scales in this problem. The first time scale,
$1/\gamma$ in the dimensional units of Eq.~(\ref{eq1}) and $1$
in the dimensionless units of Eq.~(\ref{eq15}), corresponds to
the characteristic exponential growth or decay of the population
in the absence of interactions ($C=0$). The second time scale is
the delay parameter $\tau$ in (\ref{eq13}), which is expressed
in units of the first time scale in our following investigation.

This equation is complemented by an initial history condition
\be
\label{eq16}
x(t) = x_0 \qquad (t\leq 0) \; ,
\ee
according to which
\be
\label{eq17}
y(t) = y_0 = a+ bx_0 \qquad (t\leq 0) \; .
\ee
Equation (\ref{eq15}) possesses two trivial solutions:
$x(t)=0$, under any initial conditions; and
\be
\label{eq18}
x(t) = x_0 \qquad (y_0 =\sgm_1\sgm_2 x_0) \; ,
\ee
occurring under the special initial conditions, given in
brackets. In the following, we will mainly focus our attention
on nontrivial solutions.

We now discuss the range of variation of the different
parameters.
\begin{itemize}
\item
The coefficient $a$, characterizing the initial resources
provided by Nature, is non-negative.

\item
The  coefficient $b$, controlling the impact of past
population on the present carrying capacity, can be
either positive or negative, depending on whether
production or destruction dominates. A known example
for $b<0$ is the destruction of habitat by humans,
associated with deforestation, reduction of biodiversity,
and climate changes \cite{35,36,37,38}. The destruction
of the global Earth ecosystem is caused by the rapid
growth of the human population, which is sometimes
compared with a pathological cancer process that could
result in the eventual extinction of the human population
\cite{39}. Another example of destructive activity is
firm mismanagement, and operational risks, which can
result in firm bankruptcy and even in a global economic
crisis, when many economic and financial institutions
are mismanaged \cite{40,41}. One more illustration is
the destruction of the economy of a country by a
corrupted government. In contrast, a positive $b$
corresponds to improved exploitation of resources
and increased productivity.

\item
The initial value $x_0$ of the dimensionless population is
positive.

\item
The initial value $y_0$ of the carrying capacity can be either
positive or negative. The standard case is, of course, $y_0>0$.
A negative value $y_0$ of the effective carrying capacity at
$t=0$ can be interpreted as describing a strongly destructive
action of the agents that occurred in the preceding time
interval $[-\tau,0]$.

\end{itemize}
Summarizing,
\be
\label{eq19}
a \geq 0 \; , \qquad  -\infty < b < \infty \; ,  \qquad x_0 > 0 \; ,
\qquad   -\infty < y_0 < \infty \;.
\ee
We restrict our investigation to non-negative dimensionless
population size $x(t) \geq 0$.

Finally, we need to discuss the signs $\sigma_1$ and
$\sigma_2$, that is, the signs of $\gamma$ and $C$. If gain
or birth (respectively, loss or death) prevails, $\gamma$ is
positive (respectively, negative). Similarly, $C$ is positive
(respectively, negative) if competition (respectively,
cooperation) dominates. Competition describes the fight of
individuals for scarce resources [1,2]. But in human, as well
as in animal societies, cooperation is often active through
feedback selection \cite{34}. Summarizing, there are thus four
possible types of societies, depending on the signs of
$\sigma_1$ and $\sigma_2$:
\begin{eqnarray}
\nonumber
\sgm_1 > 0  \quad \& \quad \sgm_2 > 0 &   \qquad
({\rm gain + competition}), \\
\nonumber
\sgm_1 > 0  \quad \& \quad \sgm_2 < 0 &   \qquad
({\rm gain + cooperation}), \\
\nonumber
\sgm_1 < 0  \quad \& \quad \sgm_2 > 0 &   \qquad
({\rm loss + competition}), \\
\label{eq20}
\sgm_1 < 0  \quad \& \quad \sgm_2 < 0 &   \qquad
({\rm loss + cooperation}).
\end{eqnarray}
We shall study each of these variants in turn in the following
sections.


\section{Scheme of stability analysis}

Before studying the different variants of the evolution equation
(\ref{eq15}) with (\ref{eq13}), it is necessary to explain the
methodology that we have used to deal with such equations. The
theory of linear delay equations and the stability of their
solutions have been described in detail in several books
\cite{10,11,12} and many articles (see Refs. \cite{42,43,44,45}).
Mao theorem \cite{46} proves that, for time lags close to zero,
nonlinear delay equations inherit some properties of the nonlinear
ordinary differential equations. Increasing the time lag can
result in novel solutions, which are principally different from
those obtained in the limit of small lags. One can even obtain
multistability \cite{47}. Thus, nonlinear delay equations are
essentially more difficult to study than ordinary differential
equations. The investigation of the solutions of nonlinear delay
equations is usually accomplished by combining analytical methods
to study the asymptotic stability of their stationary points,
together with the direct numerical solution of these equations
\cite{10,11,12,48}.

The study of the asymptotic stability of the stationary fixed
points of the nonlinear delay equation (\ref{eq15}) with
(\ref{eq13}) is performed using the general Lyapunov stability
analysis as follows. If stationary solutions $x^*$ exist, they
satisfy the equation
\be
\label{eq21}
\sgm_1 x^* \; - \; \frac{\sgm_2(x^*)^2}{a+bx^*} \; =
\; 0 \; .
\ee
This gives two fixed points
\be
\label{eq22}
x_1^* = 0 \; , \qquad x_2^* = \frac{a\sgm_1}{\sgm_2-b} 
\ee
that are assumed to be non-negative. Considering a small deviation
from these fixed points given by (\ref{eq22}),
\be
\label{eq23}
x_j(t) \simeq x_j^* + \dlt x_j(t) \qquad (j=1,2) \; ,
\ee
the linearized version of Eq. (\ref{eq15}) reads
\be
\label{eq24}
\frac{d}{dt} \; \dlt x_j(t) = C_j \dlt x_j(t) +
D_j \dlt x_j(t-\tau) \; ,
\ee
in which
\be
\label{eq25}
C_j \equiv \sgm_1 \; - \; \frac{2\sgm_2 x_j^*}{a+bx_j^*} \; ,
\qquad D_j \equiv b\sgm_2 \left ( \frac{x_j^*}{a+bx_j^*}
\right )^2 \; .
\ee
For the first fixed point, this yields
\be
C_1  = \sgm_1 \; , \qquad D_1 = 0 \; ,
\ee
while for the second point, this gives
\be
C_2 = \sgm_1\; \frac{b(\sgm_1-1)-\sgm_2}{b(\sgm_1-1)+\sgm_2} \; ,
\qquad D_2 = \frac{b\sgm_2}{[b(\sgm_1-1)+\sgm_2]^2} \; .
\ee
Looking for solutions of Eq. (\ref{eq23}) in the form
$ \dlt x_j(t) \propto e^{-\lambda_j t}$, we get the following
equation for the characteristic exponent $\lbd_j $:
\be
\label{eq26}
\lbd_j = C_j + D_j e^{-\lbd_j \tau} \; .
\ee
Introducing the variables
\be
\label{eq27}
W \equiv (\lbd_j - C_j)\tau \; ,
\ee
and
\be
\label{eq28}
z \equiv \tau D_j e^{-C_j\tau}
\ee
transforms Eq. (\ref{eq26}) into the equation
\be
\label{eq29}
W e^W = z \; .
\ee
The solution to this equation, in terms of the variable $W$
as a function of the variable $z$, defines the Lambert function
$W(z)$. Denoting
\be
{\hat W}(\tau) = W(z(\tau)) = W\left( \tau D_j e^{-C_j\tau}
\right)~,
\ee
allows us to obtain the solution of (\ref{eq26}) as
\be
\label{eq30}
\lbd_j = C_j + \frac{{\hat W}(\tau)}{\tau} \; .
\ee
A fixed point is stable when ${\rm Re}\;\lambda_j<0$; it is neutrally
stable when ${\rm Re}\;\lambda_j=0$, which usually defines a center; and
it is unstable if ${\rm Re}\;\lambda_j>0$.

It is necessary to keep in mind that the Lyapunov stability
analysis for a nonlinear delay equation only gives sufficient
conditions on the domain of parameters inside which the
stationary solutions are stable. According to the Mao's theorem
\cite{46}, a nonlinear delay equation possesses the same stable
fixed points as the related nonlinear ordinary differential
equation, but only in the vicinity of zero delay time.
Increasing the delay time can result in new solutions, which
are unrelated to those of the associated ordinary, non-delayed,
equation. In addition, when there are no fixed points, there
can arise different types of solutions under varying delay
time. Therefore, for delay equations, the stability analysis
has to be complemented by detailed numerical investigation. In the
following sections, we will first apply the above stability
analysis to the differential delay equation (\ref{eq15})  with
(\ref{eq13}), for the four different regimes (\ref{eq20}).
We will then complement it with a thorough numerical
investigation of the trajectories. In particular, our goal
is to present an exhaustive classification of all qualitatively
different types of solutions of Eq. (\ref{eq15})  with (\ref{eq13})
for the whole possible ranges of the parameters $a,b$, and $\tau$.


\section{Prevailing gain and competition ($\sgm_1 > 0,
\sgm_2 > 0$) \label{thnotbw}}

\subsection{General analysis}

When gain (birth) prevails over loss (death) and competition
prevails over cooperation, this corresponds to the first line
in the classification (\ref{eq20}). Then Eq. (\ref{eq15})
translates into
\be
\label{eq31}
\frac{dx(t)}{dt} = x(t) \; - \;
\frac{x^2(t)}{a+b x(t-\tau)} \; .
\ee
At the initial stage for $t<\tau$, for which $x(t-\tau)= x_0$, Eq.
(\ref{eq31}), is explicitly solvable, giving
\be
\label{eq32}
x(t) = \frac{x_0y_0e^t}{y_0+x_0(e^t-1)}  \qquad (t < \tau) \; ,
\ee
where $y_0$ is defined in Eq. (\ref{eq17}). However, the following
evolution of $x$ for $t>\tau$ cannot be described analytically.

In the general case, there are two stationary solutions
\be
\label{eq33}
x_1^*=0 \; , \qquad x_2^* =\frac{a}{1-b} \; .
\ee
The first of them is unstable for any $a>0$ and any $b$, and all
$\tau>0$. The second fixed point $x_2^*$ is stable in one of the
regions, when either
\be
\label{eq34}
a > 0 \; , \qquad -1 < b < 1 \; , \qquad  \tau \geq 0 \; ,
\ee
or
\be
\label{eq35}
a = 0 \; , \qquad 0 < b < 1 \; , \qquad \tau \geq 0 \; ,
\ee
or
\be
\label{eq36}
a > 0 \; , \qquad b < -1 \; , \qquad \tau <\tau_0 \; ,
\ee
where
\be
\label{eq37}
\tau_0 \equiv \frac{1}{\sqrt{b^2-1}} \;
\arccos \left ( \frac{1}{b} \right ) \; .
\ee
The point $x_2^*$ becomes a stable center (associated with
a vanishing  Lyapunov exponent $\lbd_2$) for
\be
\label{eq38}
a > 0 \; , \qquad b < -1 \; , \qquad \tau=\tau_0 \; .
\ee
The value $\tau_0$ diverges, if $b \nearrow -1$, as
$$
\tau_0 \simeq \frac{\pi}{\sqrt{2(|b|-1)} } \qquad
(b\nearrow -1) \; .
$$
Varying the system parameters yields the different solutions,
which we analyze successively.

\subsection{Punctuated unlimited growth}

When the carrying capacity increases, due to the intensive
creative activity of the agents forming the system, which
corresponds to the parameters
\be
\label{eq39}
a \geq 0 \; , \qquad b \geq 1 \; , \qquad \tau\geq 0 \; ,
\ee
then $x_0<y_0$ and the fixed point $x_2^*$ does not exist.
The function $x(t)$ grows by steps of duration $\simeq \tau$,
tending to infinity as time  increases to infinity. Figures
1, 2, and 3 demonstrate the behavior of $x=x(t)$ as a function
of time for different values of the parameters $a$ (Fig. 1),
$b$ (Fig. 2), and the delay time $\tau$ (Fig. 3). Different
initial conditions of $x_0$ result in the shift of the curves,
as is shown in Fig. 4. The evolution goes through a succession
of stages where $x$ is practically constant, which are interrupted
by periods of fast growth. To show that, on average, the growth
is exponential, we present in Fig. 5 the dependence of $\ln x(t)$
for a long time interval (long compared with $\tau$).

\subsection{Punctuated growth to a stationary level}

For a lower creative activity (quantified by $b$) of the
population affecting the effective carrying capacity, i.e.,
for
\be
\label{eq40}
a > (1-b) x_0 \; , \qquad 0 \leq b < 1 \; , \qquad
\tau\geq 0 \; ,
\ee
which implies that
$$
x_0 < y_0 < x_2^* \; ,
$$
the value of $x(t)$ monotonically grows to the stationary
solution $x_2^*$, as is shown in Figs. 6, 7, and 8 for the
varying parameters $a$ (Fig. 6), $b$ (Fig. 7), and $x_0$
(Fig. 8).

\subsection{Punctuated decay to a stationary level}

When the pre-existing carrying capacity $a$ is smaller than in
the previous cases and the creation coefficient $b$  is not too
high, so that
\be
\label{eq41}
0 \leq a < (1-b)x_0 \; , \qquad 0 \leq b < 1 \; , \qquad
\tau \geq 0 \; ,
\ee
which means that
$$
x_0 > x_2^* > y_0 > 0 \; ,
$$
then $x(t)$ monotonically decays to the stationary solution
$x_2^*$, as is shown in Fig. 9 for different parameters $b$.

\subsection{Punctuated alternation to a stationary level}

When the initial capacity $a$ is large, but the agent activity
is destructive, with the parameters
\be
\label{eq42}
a > |b| x_0 \; , \qquad -1 \leq b < 0 \; , \qquad
\tau \geq 0 \; ,
\ee
there are two subcases. If
$$
a > (1 +|b|) x_0 \; ,
$$
so that
$$
x_0 < x_2^* < y_0 \; ,
$$
then $x(t)$ grows initially. And if
$$
|b| x_0 < a < (1 + |b| ) x_0 \; ,
$$
so that
$$
x_0 > x_2^* > y_0 > 0 \; ,
$$
then $x(t)$ decreases initially. However, the following
behavior in both these subcases is similar: $x(t)$ tends
to the stationary solution $x_2^*$ through a sequence of
up and down alternations, as shown in Fig. 10.

\subsection{Oscillatory approach to a stationary level}

If the capacity is large and the destructive activity is
rather strong, such that
\be
\label{eq43}
a > |b| x_0 \; , \qquad b < -1 \; , \qquad
\tau < \tau_0 \; ,
\ee
where $\tau_0$ is given by Eq. (\ref{eq37}), there are
again two subcases, when $x(t)$ either increases or
decays initially. But the following behavior for both
these subcases is again similar: $x(t)$ tends towards
the focus $x_2^*$ by oscillating around it. Contrary to
the previous case 4.5, here the stagnation stages are
practically absent, so that the overall evolution is
purely oscillatory, with a decaying amplitude of
oscillations, as shown in Fig. 11.

\subsection{Everlasting nondecaying oscillations}

With the parameters $a$ and $b$ as in the previous case, but
with the time lag being exactly equal to $\tau_0$ given by
(\ref{eq37}), that is, when
\be
\label{eq44}
a > |b| x_0 \; , \qquad b < -1 \; , \qquad \tau=\tau_0 \; ,
\ee
then $x(t)$ oscillates around the center $x_2^*$ without
decaying, as shown in Fig. 12. At the initial time, $x$ can
either increase or decrease, as in the previous cases. But,
it will rapidly set into a stationary oscillatory behavior
without attenuation.

\subsection{Punctuated alternation to finite-time death}

The fact that the behavior of the system depends sensitively
on the time lag $\tau$ is well exemplified by the regime in
which the values of $a$ and $b$ are the same as in regime 4.7,
but the lag becomes longer, so that
\be
\label{eq45}
a > | b| x_0 \; , \qquad b < -1 \; , \qquad \tau > \tau_0 \; .
\ee
In this regime, $x(t)$ alternates between upward and downward
jumps, with increasing amplitude, until it hits the zero level
at a finite death time $t_d$ defined by the equation
\be
\label{eq46}
a + b x(t_d -\tau ) = 0 \; ,
\ee
at which time the rate of decay becomes minus infinity. As in
the previous cases, depending on whether $x_0<y_0$ or $x_0>y_0$,
the initial motion can be either up or down, respectively. But
the following behavior follows a similar path, with $x(t)$
always going to zero in finite time, as shown in Fig. 13. The
abrupt fall of the population $x(t)$ to zero can be interpreted
as a {\it mass extinction}, as has occurred several times for
species on the Earth [49-54]. Here, as in all previous cases,
the considered parameters are such that the initial carrying
capacity is positive, $y_0>0$. The effect of mass extinction
in the present example is caused by the intensive destructive
activity ($b<-1$) of the agents composing the system. This is
an example of total collapse caused by {\it the destruction of
habitat}.

\subsection{Growth to a fixed finite-time singularity}

Another example of catastrophic behavior happens when the
initial carrying capacity is negative ($y_0<0$). This occurs
when the habitat has been destroyed in the preceeding time
interval $[-\tau,0]$ and the destruction goes on for $t>0$.
For the set of parameters
\be
\label{eq47}
a < |b| x_0 \; , \qquad b <  0 \; , \qquad \tau \geq t_c \; ,
\ee
with a sufficiently long time lag $\tau$, the function $x(t)$,
solution of Eq. (\ref{eq32}), diverges at the singularity time
$t_c$, given by the expression
\be
\label{eq48}
t_c = \ln \left ( 1 \; - \; \frac{y_0}{x_0} \right ) \; .
\ee
The divergence is hyperbolic, i.e., in the vicinity of $t_c$,
\be
x(t) \simeq \frac{y_0}{t_c-t} \qquad ( t \ra t_c - 0 ) \; .
\label{hwgkw}
\ee
For the parameters (\ref{eq47}), the singularity always
occurs at the critical time (\ref{eq48}) determined by the
values of $x_0$ and $y_0$, independently on the delay time
$\tau$ as long as $\tau$ is larger than $t_c$.

\subsection{Growth to a moving finite-time singularity}

When the delay time $\tau$ is smaller than the singularity
time $t_c$ given by Eq.~\ref{eq48}, with the following
parameters
\be
\label{eq49}
a < |b|x_0 \; , \qquad b < 0 \; , \qquad
\tau_c < \tau \leq t_c \; ,
\ee
then the critical lag $\tau_c$, for the given parameters,
can only be determined numerically. In this regime, $x(t)$
grows without bound and reaches infinity in finite time at
a moving singularity time $t_c^*\geq t_c$ which is a
function of $\tau$. We find that  $t_c^*$ goes to infinity
as $\tau$ decreases to $\tau_c$. The dependence of the
singularity time $t_c^*$ as a function of $\tau$ is
presented in Fig. 14.

While the model does not describe what happens beyond the
singularity, the catastrophic divergence of $x(t)$ can be
interpreted as a diagnostic of a transition to another
state or to a different regime in which other mechanisms
become dominant. In analogy with the divergences occurring 
at the critical points of phase transitions of many-body 
systems \cite{55,56,57}, it is natural to interpret the 
critical points as periods of transitions to new regimes. 
Ref. \cite{58} has reviewed several examples of the application 
and interpretation of the occurrence of finite-time 
singularities in the dynamics of the world population, 
economics, and finance.

\subsection{Exponential growth to infinity}

As the delay time $\tau$ becomes smaller than the threshold
value $\tau_c$ defined in the previous section, i.e., for
the following parameters
\be
\label{eq50}
a < |b| x_0 \; , \qquad b < 0\; , \qquad
0 < \tau \leq \tau_c \; ,
\ee
the finite-time singularity does not exist anymore. The
function $x(t)$ exhibits a simple unbounded exponential
growth to infinity, as time tends to infinity.

The exact limit of a zero time delay $\tau=0$ is not
included in this regime. When $\tau$ is exactly zero,
the exponential growth regime is replaced abruptly into
the regime of subsection 4.9, with a fixed-time singularity.

Figure 15 demonstrates the change of behavior of $\ln x(t)$
as a function of time for different values of the delay
time $\tau$, for fixed parameters $a$ and $b$. In this
regime, the variable $y(t)$ tends to minus infinity with
increasing time, and it becomes difficult to interpret it
as an effective carrying capacity. Rather, this regime with
negative $b$ expresses the existence of a positive feedback
provided by the cooperation between agents of the system.


\section{Prevailing gain and cooperation ($\sgm_1 > 0,
\sgm_2 < 0$)}

\subsection{General analysis}

When gain (birth) and cooperation prevail (second line of
classification (\ref{eq20})), Eq. (\ref{eq15}) becomes
\be
\label{eq51}
\frac{dx(t)}{dt} = x(t) + \frac{x^2(t)}{a+bx(t-\tau)} \; .
\ee
The same history $x(t) = x_0$ for $t\leq 0$ as in (\ref{eq16})
is assumed. For $t<\tau$, for which $x(t-\tau)=x_0$, the
solution is
$$
x(t) = \frac{x_0y_0 e^t}{y_0-x_0(e^t-1)} \qquad
(t < \tau ) \; ,
$$
with $y_0$ given by Eq. (\ref{eq17}).

Equation (\ref{eq51}) possesses two fixed points
\be
\label{eq52}
x_1^* = 0 \; , \qquad x_2^* = -\; \frac{a}{b+1} \; .
\ee
The second fixed point $x_2^*$ is positive for $b<-1$. The
stability analysis, performed following the methodology
explained in Section 3, shows that the first point $x_1^*$
is always unstable. The second fixed point $x_2^*$ can be
stable, while non-negative, only for
\be
\label{eq53}
a = 0 \; , \qquad - 1 < b < 0 \; , \qquad \tau \geq 0 \; .
\ee
For $a \to 0$, it merges with the first fixed point. The
full analysis yields the following different types of
solutions.

\subsection{Growth to a fixed finite-time singularity}

For the parameters
\be
\label{eq54}
a > - bx_0 \; , \qquad \forall b \; , \qquad
\tau \geq t_c \; ,
\ee
when $y_0>0$, the solution is monotonically increasing and
becomes singular at the finite catastrophe time
\be
\label{eq55}
t_c = \ln \left ( 1 + \frac{y_0}{x_0} \right ) \; ,
\ee
which does not depend on the delay time $\tau$,
similarly to the behavior in Sec. 4.9. The occurrence
of the singularity, in the presence of the positive
initial carrying capacity and positive production
coefficient $b$, shows that the simultaneous gain and
cooperation is not sustainable, since the system diverges
in finite time. This paradox, that initial positive carrying
capacity and ever increasing carrying capacity, intrinsically
associated with a positive feedback, leads to a run-away,
has been analyzed in detail in Ref. \cite{58} in another
context. To preserve stability, one would need that either
the gain has to turn into loss or cooperation to be replaced
by competition.

\subsection{Growth to a moving finite-time singularity}

For smaller delay time, when either
\be
\label{eq56}
a \geq 0 \; , \qquad b > 0\; , \qquad
\tau_c < \tau < t_c \; ,
\ee
or
\be
\label{eq57}
a > | b| x_0 \; , \qquad  b < 0 \; , \qquad
0 < \tau \leq t_c \; ,
\ee
the time of the singularity becomes dependent on the lag.
The singularity occurs at $t_c^*>t_c$, if $b>0$ and at
$t_c^*<t_c$, if $b<0$. The divergence is hyperbolic as
in expression (\ref{hwgkw}). Keeping the parameters $a$
and $b$ fixed, but increasing the delay time $\tau$, moves
the singularity time $t_c^*$ to the left towards $t_c$ for
$b>0$, while $t_c^*$ moves to the right again towards $t_c$
for $b<0$.

\subsection{Exponential growth to infinity}

Under the conditions
\be
\label{eq58}
a \geq 0 \; , \qquad b > 0 \; , \qquad
0 < \tau \leq \tau_c \; ,
\ee
the function $x(t)$ grows exponentially without bounds. The
behavior of $x(t)$ is analogous to that  of Sec. 4.11. The
growth of $x(t)$ is continuous, following the increase of
the effective carrying capacity associated with the positive
production coefficient $b$. Contrary to the previous cases
of Sections 5.2 and 5.3, there is no finite-time catastrophe
as the shorter delay time allows for a better matching the
carrying capacity and population size.

\subsection{Punctuated unlimited growth}

When the initial carrying capacity is negative, $y_0<0$,
having been destroyed in the preceeding time interval
$[-\tau,0]$, and the parameters are
\be
\label{eq59}
a \geq 0 \; , \qquad b < -1 - \; \frac{a}{x_0} \; ,
\qquad \tau \geq 0 \; ,
\ee
$x(t)$ follows a punctuated unlimited growth, as in Sec. 4.2.
In this regime, $y(t)$ remains negative and goes to $-\infty$
at large times. Thus, $y(t)\ra-\infty$ is difficult to
interpret as an effective carrying capacity.

\subsection{Punctuated decay to finite-time death}

For the parameters
\be
\label{eq60}
a > 0 \; , \qquad -1 - \; \frac{a}{x_0} < b < 0 \; ,
\qquad \tau \geq 0 \; ,
\ee
there exists a time $t_d$, defined by Eq. (\ref{eq46}),
when $x(t)$ sharply drops to zero, as shown in Fig. 16.
This is the point of mass extinction. While the decay
of $x(t)$ is a finite succession of plateaus and drops,
contrary to the case of  Sec. 4.8, there are no
oscillations, but just a monotonic decay to death.
The disappearance of $x(t)$ in this regime is due to
the destructive activity ($b<0$) aggravated by the
 prevailing birth (gain) and cooperation.

\subsection{Punctuated decay to zero}

For the parameters as in (\ref{eq53}), that is,
\be
a=0,-1<b<0,\tau\geq 0 ,
\ee
$x(t)$ decays to zero as time
tends to infinity in a punctuated fashion following a
succession of plateaus followed by sharp drops. This
behavior is illustrated in Fig. 17 for different values
of the parameters. As in section 5.5, the decay of $x(t)$
is due to the destructive activity ($b<0$) aggravated by
the prevailing birth (gain) and cooperation.


\section{Prevailing loss and competition ($\sgm_1 < 0,
\sgm_2 > 0$)}

\subsection{General analysis}

Let us now consider the regime corresponding to the third
line of classification (\ref{eq20}). In this case, Eq.
(\ref{eq15}) takes the form
\be
\label{eq61}
\frac{dx(t)}{dt} = - x(t)  -\; \frac{x^2(t)}{a+bx(t-\tau)} \; .
\ee
The initial history is the same as in Eq. (\ref{eq16}). In the
time interval $t<\tau$, the solution reads
\be
\label{eq62}
x(t) = \frac{x_0y_0e^{-t}}{y_0+x_0(1-e^{-t})} \qquad
(t < \tau ) \; ,
\ee
with $y_0$ given by Eq. (\ref{eq17}).

There are two fixed points
\be
\label{eq63}
x_1^* = 0 \; , \qquad x_2^* = -\; \frac{a}{1+b}~.
\ee
The second fixed point $x_2^*$ is relevant only
when it is non-negative.

The stability analysis,
complemented by numerical investigations, shows the following
properties. The first stationary solution $x_1^*=0$ is stable
for the parameters
\be
\label{eq64}
a \neq 0 \; , \qquad \forall b \; , \qquad \tau \geq 0\; .
\ee
The second stationary solution $x_2\neq 0$ is stable, while being
positive, for
\be
\label{eq65}
a > 0 \; , \qquad b < -1 \; , \qquad \tau < \tau_0 \; ,
\ee
where
\be
\label{eq66}
\tau_0 \equiv \frac{1}{\sqrt{b^2-1}} \; \arccos\left ( -\;
\frac{1}{b} \right ) \; .
\ee

A distinct regime occurs for $a = 0$, for which the two fixed
points merge together ($x_1^*=x_2^*=0$). This double fixed point
$0$ is stable when either
\be
\label{eq67}
a = 0 \; , \qquad b < -1 \; , \qquad \tau \geq 0
\ee
or
\be
\label{eq68}
a = 0 \; , \qquad b > 0 \; , \qquad \tau \geq 0 \; .
\ee

The domains (\ref{eq64}) and (\ref{eq65}) overlap, indicating
the existence of bistability, i.e., both stationary solutions
(\ref{eq63}) are stable simultaneously for $a>0$ and $b<-1$.
For these parameters, the solution $x(t)$ tends to one of these
fixed points depending on whether the initial condition $x_0$
falls in the domain of attraction of the first or second fixed
point. The basin of attraction of the fixed point $x_1^*$
(respectively, $x_2^*$)  is defined by the condition $x_0< x_2^*$
(respectively, $x_0>x_2^*$).

\subsection{Monotonic decay to zero}

For any positive starting carrying capacity $y_0>0$, when
\be
\label{eq69}
a \geq 0 \; , \qquad b > - \; \frac{a}{x_0} \; , \qquad
\tau \geq 0 \; ,
\ee
the solutions to Eq. (\ref{eq61}) always monotonically decay to
zero, as is shown in Fig. 18. The case (\ref{eq68}) is included
here. The same behaviour occurs under conditions (\ref{eq67}),
though then $y_0$ is negative. The meaning of such a decay is
evident: Prevailing loss and competition do not favor the system
development.

\subsection{Oscillatory approach to a stationary level}

The situation is much more ramified, when $y_0<0$. If
$y_0<-x_0$, so that
\be
\label{eq70}
a > 0 \; , \qquad b < -1 - \; \frac{a}{x_0} \; , \qquad
0 < \tau < \tau_0 \; ,
\ee
where $\tau_0$ is given by Eq. (\ref{eq66}), then the solution
$x(t)$ oscillates around the focus $x_2^*$, tending to it as
time increases. This behaviour is illustrated by Fig. 19.

\subsection{Everlasting nondecaying oscillations}

In the region of the parameters
\be
\label{eq71}
a > 0 \; , \qquad b < -1 - \; \frac{a}{x_0} \; , \qquad
\tau_0 \leq \tau < \tau_1 \; ,
\ee
where $\tau_1$ depends on the values of $a$ and $b$, the
solution $x(t)$ oscillates without decay, as is presented
in Fig. 20. The amplitude of the oscillations increases
with increasing time lag $\tau$. Note that the period of
the oscillations is much longer than the time lag. Though
this case looks similar to that studied in Sec. 4.6, there
is a principal difference. Here, the oscillations persist
not just for one particular lag $\tau_0$, but in the whole
interval of lags, given in Eq. (\ref{eq71}).

\subsection{Punctuated growth to a moving finite-time
singularity}

With the same range of the parameters $a$ and $b$, as
in Eq. (\ref{eq71}), but with the larger delay time in the
interval
\be
\label{eq72}
\tau_1 \leq \tau < \tau_2 \; ,
\ee
when $y_0<-x_0$, a finite-time singularity occurs at time
$t_c^*$, defined by the equation
\be
\label{eq73}
a + bx (t_c^* -\tau ) = 0 \; ,
\ee
where $x(t)$ diverges hyperbolically as in Eq.~(\ref{hwgkw}).
The second characteristic delay time $\tau_2$ depends also on
the values of $a$ and $b$. The difference from the divergences
of Sec. 4.10, depicted in Fig. 15, is that in the present case the
function $x(t)$ first decreases with time before accelerating
towards the singularity. In Fig. 21, we observe that there can
be several punctuated phases, the first decay followed by growth,
followed by decay, followed by the final acceleration to infinity
in finite time.

For  $\tau > \tau_2$, the divergence is replaced by a monotonic
decay to zero, as in Sec. 6.2.

The cases of Secs. 6.3, 6.4, and 6.5 illustrate how, being in
the same range of the parameters $a$ and $b$, but merely varying
the time lag $\tau$, can result in a principally different
behaviour of the solution $x(t)$.

\subsection{Up-down approach to a stationary level}

The cases of Secs. 6.3, 6.4, and 6.5 correspond to $y_0<-x_0$.
We now consider the case
\be
\label{eq74}
- x_0 < y_0 < 0 \; ,
\ee
for which the behavior $x(t)$ depends on the relative value of
$\tau$ compared with a characteristic $\tau_c$, which can be
defined only numerically, being such that
\be
\label{eq75}
\tau_c < \tau_0 < t_c \; ,
\ee
where $\tau_0$ is given by Eq.~\ref{eq37} and
\be
\label{eq76}
t_c \equiv - \ln \left ( 1 + \frac{y_0}{x_0} \right ) \; .
\ee
With a good approximation, $\tau_c$ is close to $\tau_0$.
For the parameters  satisfying the conditions
\be
\label{eq77}
a > 0 \; , \qquad -1 - \; \frac{a}{x_0} < b < -1 \; , \qquad
0 < \tau < \tau_c \; ,
\ee
the solution $x(t)$ first increases sharply before
decaying to the stationary point $x_2^*$, as shown in Fig. 22.

\subsection{Growth to a finite-time singularity}

There are several cases of infinite growth occurring in finite
time, which are similar to one of the cases considered above,
although they happen under quite different conditions. For the
parameters
\be
\label{eq78}
a \geq 0 \; , \qquad -1 - \; \frac{a}{x_0} < b < -\;
\frac{a}{x_0} \; , \qquad \tau > t_c \; ,
\ee
with $t_c$ given by Eq. (\ref{eq76}), the divergence occurs at
time $t_c$, analogously to the behavior documented in Sec. 5.2.

When the delay time $\tau$ becomes smaller than $t_c$, in the
region of the parameters
\be
\label{eq79}
a \geq 0 \; , \qquad -1 - \; \frac{a}{x_0} < b < - \;
\frac{a}{x_0} \; , \qquad \tau_c < \tau \leq t_c \; ,
\ee
there exists a critical time $t_c^*$ which is a function of
$a, b$ and $\tau$, at which $x(t)$ diverges hyperbolically
as in Sections 5.3 or 6.5.

\subsection{Exponential growth to infinity}

For time delays $\tau$ smaller than $\tau_c$ and the parameters
\be
\label{eq80}
a \geq 0 \; , \qquad -1 < b < - \; \frac{a}{x_0} \; , \qquad
0 < \tau \leq \tau_c \; ,
\ee
the point of singularity moves to infinity, and $x(t)$ exhibits
an exponential growth with time, as in Secs. 4.11 or 5.4.


\section{Prevailing loss and cooperation ($\sgm_1 < 0,
\sgm_2 < 0$)}

\subsection{General analysis}

If loss (death) and cooperation prevail, this corresponds
to the fourth line of classification (\ref{eq20}). Then,
Eq. (\ref{eq15}) becomes
\be
\label{eq81}
\frac{dx(t)}{dt} = - x(t) + \frac{x^2(t)}{a+bx(t-\tau)} \; .
\ee
As everywhere above, the same history condition (\ref{eq16})
is assumed. As above, there exists a first regime for $t<\tau$,
for which the solution can be written explicitly as
\be
\label{eq82}
x(t) = \frac{x_0y_0e^{-t}}{y_0-x_0(1-e^{-t})} \; \qquad
(t < \tau ) \; ,
\ee
where $y_0$ is given by Eq. (\ref{eq17}).

Equation (\ref{eq81}) possesses two stationary points
\be
\label{eq83}
x_1^* = 0 \; , \qquad x_2^* = \frac{a}{1-b} \; .
\ee
The second fixed point $x_2^*$ is non-negative for either
$a>0$ and $b<1$ or $a\equiv 0$. The stability analysis of
Sec. 3 shows that the first fixed point $x_1^*$ is stable
for all nonzero $a$, any $b$, and any $\tau$, as in Eq.
(\ref{eq64}). In contrast, the second fixed point $x_2^*$
is unstable for any $a>0$ and $b<1$ for which it is positive.

The special case $a\equiv 0$ is such that the two fixed points
merge together: $x_1^*=x_2^*\equiv x^*=0$. Then the Lyapunov
stability analysis of Sec. 3 indicates that $x^*$ is unstable
for $0<b<1$. It is stable, when either
\be
\label{eq84}
a = 0 \; , \qquad b < 0 \; , \qquad \tau \geq 0\; ,
\ee
or
\be
\label{eq85}
a = 0 \; , \qquad b > 3 \; , \qquad \tau \geq 0 \; ,
\ee
or
\be
\label{eq86}
a = 0 \; , \qquad 1 < b < 3 \; , \qquad \tau < \tau_0 \; ,
\ee
where
\be
\label{eq87}
\tau_0 \equiv \frac{b}{\sqrt{(b-1)(3-b)}} \;
\arccos (2 - b) \qquad (1 < b < 3) \; .
\ee

It is worth stressing that these results follow from the
stability analysis of the linearized delay equation. As
has been emphasized in Sec. 3, the Lyapunov analysis of
the asymptotic stability for the delay equations gives
only sufficient conditions of stability. The determination
of the actual region of parameters, for which the stationary
solutions of a delay equation are stable, requires to
complement the Lyapunov analysis by detailed numerical
checks. This has been done everywhere above.

To stress the necessity of accomplishing detailed numerical
calculations for delay equations, let us consider the present
case that provides a good illustration of such a necessity.
The Lyapunov analysis for the fixed point $x^*=0$, when
$1<b<3$, indicates that this point is stable for $\tau<\tau_0$,
as expressed by conditions (\ref{eq86}). However, from numerical
calculations, it follows that the actual region of stability is
larger, and extends to all $\tau>0$. In order to illustrate this
difference, we present in Fig. 23, for the case of $a=0$, the
solutions to the exact delay equation (\ref{eq81}), compared
with its linearized variant. The linearized approximation
exhibits an unstable behavior, while the solution to the full
Eq. (\ref{eq81}) is stable.

In this way, we have determined that the stationary solution
$x^*=0$ is stable for all non-negative $a$, any $b$, and all
$\tau$.

\subsection{Monotonic decay to zero}

When $x_0<y_0$, and the parameters are
\be
\label{eq88}
a \geq 0 \; , \qquad b > 1 - \; \frac{a}{x_0} \; ,
\qquad \tau \geq 0 \; ,
\ee
the function $x(t)$ monotonically decays to zero. This
behavior is similar to that of Sec. 6.2, with the difference
that, for some parameters, the decaying solution does not
have a constant convexity, as is demonstrated in Fig. 24.
A similar decay to zero occurs for $y_0<0$, with the parameters
\be
\label{eq89}
a = 0 \; , \qquad b < 0 \; , \qquad \tau \geq 0 \; .
\ee
This decaying behavior is caused by the prevailing loss.

\subsection{Growth to a fixed finite-time singularity}

For $0<y_0<x_0$ and
\be
\label{eq90}
a > 0 \; , \qquad -\; \frac{a}{x_0} < b < 1 - \;
\frac{a}{x_0} \; , \qquad \tau \geq t_c \; ,
\ee
where
\be
\label{eq91}
t_c \equiv - \ln \left ( 1  - \; \frac{y_0}{x_0}
\right ) \; ,
\ee
the solution $x(t)$ diverges hyperbolically at $t_c$. The
finite singularity time $t_c$ is independent of the delay
time $\tau$ as long as the later remains larger than $t_c$.

\subsection{Growth to a moving finite-time singularity}

For $\tau< t_c$, the catastrophic divergence occurs at a
singularity time $t_c^*$, whose value depends on $\tau$. For
the parameters
\be
\label{eq92}
a >  0\; , \qquad - \; \frac{a}{x_0} < b \leq 0 \; ,
\qquad 0 < \tau < t_c \; ,
\ee
the divergence occurs at $t_c^*<t_c$, which moves to the
right towards $t_c$, as $\tau$ increases to $t_c$. In
contrast, for the parameters
\be
\label{eq93}
a > 0  \; , \qquad 0 < b < 1  -\; \frac{a}{x_0} \; ,
\qquad \tau_c < \tau < t_c \; ,
\ee
where $\tau_c$ depends on the parameters $a$ and $b$, the
divergence occurs at a point $t_c^*>t_c$, which moves to
the left towards $t_c$, as $\tau$ increases. This behavior
is analogous to that described in Sec. 5.3.

\subsection{Exponential growth to infinity}

For the parameters $a$ and $b$ as defined in Eq. (\ref{eq93}),
but for smaller time lags, such that
\be
\label{94}
0 < \tau \leq \tau_c \; ,
\ee
the solution $x(t)$ to Eq. (\ref{eq81}) grows exponentially
at large times, similarly to the regimes documented in Secs.
5.4 and 6.8.

\subsection{Monotonic decay to finite-time death}

For $y_0<0$, and for the parameters
\be
\label{eq95}
a > 0 \; , \qquad b < - \; \frac{a}{x_0} \; , \qquad
\tau \geq 0 \; ,
\ee
the solution $x(t)$ decays monotonically to zero in finite
time, reaching zero at time $t_d$, defined by the equation
$a+bx(t_d-\tau )=0$. When $t$ tends to $t_d$ from below, $x(t)$
sharply drops to zero, as shown in Fig. 25. This behavior
differs from the finite-time death regime described in Sec. 4.8
by the absence of punctuated alternations. And it also differs
from the finite-time death regime of Sec. 5.6 by the absence of
punctuated steps. Here, death in the present regime is due to
the destructive activity of agents under the prevailing loss
term in Eq. (\ref{eq81}).


\section{Conclusion}

We have suggested a new variant of the logistic-type equation,
in which the carrying capacity consists of two terms. One
of them corresponds to a fixed carrying capacity, provided
by Nature. And another term is the carrying capacity created
(or destroyed) by the activity of the agents composing the
considered society. The created (or destroyed) capacity is
naturally delayed, as far as any creation/destruction
requires some time. We take into account both the creative
but also destructive impact of the agents on the carrying
capacity. This is quantified by a coefficient $b$ of activity
which is positive for creation and negative for destruction.

Four different situations have been analyzed, depending on
the signs of the two terms entering our delay logistic
equation. We have taken into account that the growth rate
is positive, when gain (birth) prevails and it is negative
if loss (death) is prevailing. The sign of the second term
in the equation is negative if the competition between agents
is stronger than the effects resulting from their cooperation.
The sign of the second term turns positive when their
cooperation is dominant.

We have carefully investigated all the possible emergent
regimes, using both analytical and numerical methods. This
has led to a complete classification of the possible types
of different solutions. It turns out that there exists a
large variety of solution types. In particular, we find
a rich and rather sensitive dependence of the structural
properties of the solutions on the value of the delay time
$\tau$. For instance, in the regime where loss and competition
are dominant, depending on the value of the initial carrying
capacity and of $\tau$, we find monotonic decay to zero,
oscillatory approach to a stationary level, sustainable
oscillations and moving finite-time singularities. This
should not be of too much surprise, since delay equations
are known to enjoy much richer properties than ordinary
differential equations. In this spirit, Kolmanovskii and
Myshkis \cite{11} provide an example of a delay-differential
equation, whose properties are as rich as those of a system
of ten coupled ordinary differential equations. We have
illustrated in different figures the main qualitative
properties of the different solutions, not repeating the
presentations of solutions with similar behavior.

The rich variety of all possible regimes is summarized in 
Figs. 26 to 29. The solution types, occurring in the most
realistic and the most complicated case of Sec.4, when gain
and competition prevail, is presented in the scheme of Fig. 26.
Respectively, Fig. 27 illustrates the admissible regimes of 
Sec. 5, when gain prevails over loss and cooperation is 
stronger than competition. Figure 28 summarizes the 
qualitatively different regimes of Sec. 6, when loss prevails 
over gain and competition is stronger than cooperation. Finally,
Fig. 29 demonstrates the variety of solution types described
in Sec. 7, when loss is larger than gain and competiton prevails 
over collaboration.

The main goal of the present paper has been the formulation of 
the novel model of evolution and the study of the mathematical and 
structural properties of its solutions. In the introduction, we 
have briefly mentioned some possible interpretations. To keep 
the paper within reasonable size limits, we defer the detailed 
discussions of possible applications to future publications. 
However, in order to connect the particular solutions, and the 
related figures, to real-life situations, we can mention several 
concrete cases, where experimental observations are summarized in 
explicit graphs.  

The experimental data for the dynamics of the world urban population 
growth, presented in Ref. \cite{15} (Fig. 7) is very close to the 
punctuated evolution of our Figs. 1 and 2. The total world population 
growth, considered back to $10^6$ years 
before present (see Figs. 1 and 2 in Ref.\cite{59}) is also reminiscent
of our Fig. 2. The data on the energy 
production and consumption, listed in several tables and figures of 
the BP Statistical Review of World Energy \cite{60}, are again very 
similar to our Figs. 1, 2, and 6. Percentage of women in economics 
\cite{61} (Figs. 1, 2, and 3) follow the ladder-type dynamics as in 
our Figs. 1 and 2. The same type of  ladder evolution is documented
for the growth of some cells in humans \cite{62} (Fig. 1) and in
insects \cite{63} (Fig. 3). The decay of human cells \cite{64} 
(Fig. 6) is analogous to that of our Figs 9 or 16. The diminishing 
woman fertility rate in European countries \cite{65} (Figs. 6 and 7)
corresponds to our Fig. 9. There are many other experimenatal data 
whose dynamics correspond to some of the solutions we have described
and whose more quantitative investigation is left for the future.

\vskip 5mm

{\bf Acknowledgement}: We acknowledge financial support 
from the ETH Competence Center "Coping with Crises in Complex 
Socio-Economic Systems" (CCSS) through ETH Research 
Grant CH1-01-08-2.



\newpage

\begin{center}

{\Large{\bf Figure Captions } }

\end{center}

{\bf Fig. 1}. Solutions $x(t)$ to Eq. (\ref{eq31}) as
functions of time for the parameters $a=2$ (solid line)
and $a=5$ (dashed-dotted line). Other parameters are:
$x_0=1$, $b=1$, and $\tau=20$. Solutions grow monotonically
by steps of duration $\simeq \tau$, and $x(t)\ra+\infty$
as $t\ra+\infty$.

\vskip 5mm

{\bf Fig. 2}. Solutions $x(t)$ to Eq. (\ref{eq31}) as
functions of time for the parameters $b=2$ (solid line)
and $b=3$ (dashed-dotted line). Other parameters are:
$a=2$, $x_0=1$, and $\tau=20$. Solutions grow monotonically
by steps  of duration $\simeq \tau$, and $x(t)\ra+\infty$
as $t\ra\infty$.

\vskip 5mm

{\bf Fig. 3}. Solutions $x(t)$ to Eq. (\ref{eq31}) as
functions of time for the parameters $\tau=10$ (solid line)
and $\tau=20$ (dashed-dotted line). Other parameters are:
$a=2$, $x_0=1$, and $b=1$. Solutions grow monotonically by
steps  of duration $\simeq \tau$, and $x(t)\ra+\infty$
as $t\ra\infty$.

\vskip 5mm

{\bf Fig. 4}. Solutions $x(t)$ to Eq. (\ref{eq31}) as
functions of time for the parameters $x_0=1$ (solid line)
and $x_0=3$ (dashed-dotted line). Other parameters are:
$a=2$, $b=1$, and $\tau=20$. Solutions grow monotonically
by steps  of duration $\simeq \tau$, and $x(t)\ra+\infty$
as $t\ra\infty$.

\vskip 5mm

{\bf Fig. 5}. Logarithm of the solutions $x(t)$ to Eq.
(\ref{eq31}) as functions of time for the parameters
$b=2$ (solid line) and $b=5$ (dashed-dotted line)
exemplifying their average long-term exponential growth.
Other parameters are: $a=2$, $x_0=1$, and $\tau=10$.

\vskip 5mm

{\bf Fig. 6}. Solutions $x(t)$ to Eq. (\ref{eq31}) as
functions of time for the parameters $a=2$ (solid line),
$a=3$ (dashed line), and $a=4$ (dashed-dotted line).
Other parameters are: $x_0=1$, $b=0.5$, and $\tau=20$.
The solutions $x(t)$ monotonically grow by steps to
their stationary points $x_2^*=a/(1-b)$ as $t\ra\infty$.
Stationary points are: $x_2^*=4$ (for solid line),
$x_2^*=6$ (for dashed line), and $x_2^*=8$ (for
dashed-dotted line).

\vskip 5mm

{\bf Fig. 7}. Solutions $x(t)$ to Eq. (\ref{eq31}) as
functions of time for the parameters $b=0.25$ (solid
line), $b=0.5$ (dashed line), and $b=0.75$ (dashed-dotted
line). Other parameters are: $a=2$, $x_0=1$, and $\tau=20$.
The solutions $x(t)$ monotonically grow by steps to their
stationary points $x_2^*=a/(1-b)$ as $t\ra\infty$.
Stationary points are: $x_2^*=8/3\approx 2.67$ (for
solid line), $x_2^*=4$ (for dashed line), and $x_2^*=8$
(for dashed-dotted line).

\vskip 5mm

{\bf Fig. 8}. Solutions $x(t)$ to Eq. (\ref{eq31}) as
functions of time for the parameters $x_0=0.5$ (solid
line), $x_0=2$ (dashed line), and $x_0=3.5$ (dashed-dotted
line). Other parameters are: $a=2$, $b=0.5$, and $\tau=20$.
The solutions $x(t)$ monotonically grow by steps to their
stationary point $x_2^*=a/(1-b)=4$ as $t\ra\infty$.

\vskip 5mm

{\bf Fig. 9}. Solutions $x(t)$ to Eq. (\ref{eq31}) as
functions of time for the parameters $b=0.1$ (solid line),
$b=0.4$ (dashed line), and $b=0.7$ (dashed-dotted line).
Other parameters are: $a=0.25$, $x_0=3$, and $\tau=20$.
The solutions $x(t)$ monotonically decrease by steps to
their stationary points $x_2^*=a/(1-b)$ as $t\ra\infty$.
Stationary points are: $x_2^*=5/18\approx 0.278$ (for solid
line), $x_2^*=5/12\approx 0.417$ (for dashed line), and
$x_2^*=5/6\approx 0.833$ (for dashed-dotted line).

\vskip 5mm

{\bf Fig. 10}. Solutions $x(t)$ to Eq. (\ref{eq31}) as
functions of time for the parameters $b=-0.66$ (solid
line) and $b=-0.33$ (dashed-dotted line). Other parameters
are: $a=3$, $x_0=1$, and $\tau=20$. The solutions $x(t)$
non-monotonically grow by steps to their stationary points
$x_2^*=a/(1-b)$ as $t\ra\infty$. The stationary points are:
$x_2^*=1.80723$ (for solid line) and $x_2^*=2.25564$
(for dashed-dotted line).

\vskip 5mm

{\bf Fig. 11}. Solution $x(t)$ to Eq. (\ref{eq31}) as a
function of time for the parameter $\tau=5<\tau_0=5.91784$.
Other parameters are: $a=3$, $x_0=1$, and $b=-1.1$. The
solution $x(t)$ tends in an oscillatory manner to its
stationary point $x_2^*=a/(1-b)=1.42857$ as $t\ra\infty$.

\vskip 5mm

{\bf Fig. 12}. Solution $x(t)$ to Eq. (\ref{eq31}) as a
function of time for the parameter $\tau=\tau_0=5.91784$.
Other parameters are: $a=2$, $x_0=1$, and $b=-1.1$. The
solution $x(t)$ oscillates around its stationary point
$x_2^*=a/(1-b)=0.952381$ as $t\ra\infty$.

\vskip 5mm

{\bf Fig. 13}. Solutions $x(t)$ to Eq. (\ref{eq31}) as
functions of time for the parameters $b=-1.25$ (solid
line) and $b=-1.75$ (dashed-dotted line). Other parameters
are: $a=3$, $x_0=1$, and $\tau=20$. The solutions $x(t)$
oscillate with increasing amplitude until time
$t_d=149.932$ (for solid line) and $t_d=105.972$ (for
dashed-dotted line) defined by the equation
$a+bx(t_d-\tau)=0$, at which
$\dot{x}(t)|_{t\ra t_d-0}=-\infty$.

\vskip 5mm

{\bf Fig. 14}. Dependence of the critical time $t_c^*(\tau)$,
where the function $x(t)$ exhibits a singularity, as a
function of the lag $\tau$. Here the parameters are: $a=0$,
$x_0=1$, and $b=-1.5$. At time $t=t_c^*$, the solution has
a singularity: $x(t)|_{t\ra t_c^*-0}=+\infty$. The instant
$t_c^*$ is defined by $a+bx(t_c^*-\tau)=0$. The time
$t_c=\ln(1-y_0/x_0)=0.916291$ is the point of singularity.

\vskip 5mm

{\bf Fig. 15}. Solutions $x(t)$ to Eq. (\ref{eq31})
as functions of time for the parameters $\tau=0.1$
(dashed-dotted line), $\tau=0.2$ (dotted line), $\tau=0.3$
(dashed line), and $\tau=0.4$ (solid line). The other
parameters are: $a=2$, $x_0=2$, and $b=-1.5$. For all
$\tau>t_c=\ln(1-y_0/x_0)=0.405465$, we have $t_c^*=t_c$.
If $\tau\leq t_c$, there exists $\tau_c\approx 0.27217$ such
that if $0<\tau\leq\tau_c$, then $x(t)$ grows exponentially
to $+\infty$. If $\tau_c<\tau\leq t_c$, then there exists a
point of singularity $t_c^*>t_c$, defined by $a+bx(t_c^*-\tau)=0$,
such that $x(t)|_{t\ra t_c^*-0}=+\infty$. When $\tau\to t_c-0$,
then $t_c^*\ra t_c+0$. The values of $t_c^*$ are respectively
$t_c^*=0.40560$ (for solid line, $\tau=0.4$) and $t_c^*=0.50993$
(for dashed line, $\tau=0.3$).

\vskip 5mm

{\bf Fig. 16}.  Solutions $x(t)$ to Eq. (\ref{eq51}) as
functions of time for the parameters $b=-1.5$ (solid line)
and $b=-1.8$ (dashed-dotted line). Other parameters are:
$a=1$, $x_0=1$, and $\tau=10$. The solutions $x(t)$ decrease
by steps until time $t_d=10.6932$ (for solid
line) and $t_d=22.0171$ (for dashed-dotted line) defined by
the equation $a+bx(t_d-\tau)=0$. At these times,
$\dot{x}(t)|_{t\ra t_d-0}=-\infty$.

\vskip 5mm

{\bf Fig. 17}. Solutions $x(t)$ to Eq. (\ref{eq51}) as
functions of time for the parameters $b=-0.25$ (solid line),
$b=-0.5$ (dashed line), and $b=-0.75$ (dashed-dotted line).
Other parameters are: $a=0$, $x_0=1$, and $\tau=20$. The
solutions $x(t)$ monotonically decrease by steps to their
stationary point $x^*=0$ as $t\ra\infty$.

\vskip 5mm

{\bf Fig. 18}. Solutions $x(t)$ to Eq. (\ref{eq61}) as
functions of time for the parameters $b=-2$ (solid line)
and $b=8$ (dashed-dotted line). Other parameters are:
$a=3$, $x_0=1$, and $\tau=10$. The solutions $x(t)$
monotonically decrease to their stationary point
$x^*=0$ as $t\ra\infty$.

\vskip 5mm

{\bf Fig. 19}. Solutions $x(t)$ to Eq. (\ref{eq61}) as
functions of time for the parameters $\tau=0.4$ (solid
line) and $\tau=0.5$ (dashed-dotted line), where
$\tau<\tau_0=0.505951$. Other parameters are: $a=1$,
$x_0=1$, and $b=-2.5$. The solutions $x(t)$ converge by
oscillating to their stationary point $x_2^*=-a/(b+1)=2/3$
as $t\ra\infty$.

\vskip 5mm

{\bf Fig. 20}. Solution $x(t)$ to Eq. (\ref{eq61}) as
a function of time for the parameter $\tau=0.605>\tau_0$,
where $\tau_0=0.505951$. Other parameters are the same as
for Fig. 19. The solution $x(t)$ exhibits sustained
oscillations with an amplitude which is an increasing
function of the delay time $\tau$ and a period much larger
than $\tau$.

\vskip 5mm

{\bf Fig. 21}. Logarithmic behavior of solutions $x(t)$
to Eq. (\ref{eq61}) as functions of time for the parameters
$\tau=0.626$ (solid line), $\tau=1.126$ (dashed line),
$\tau=1.626$ (dotted line), and $\tau=2.126$ (dashed-dotted
line), where $\tau_1<\tau<\tau_2$. Other parameters are the
same as for Fig. 19. There exist points of singularity
$t_c^*$, defined by $a+bx(t_c^*-\tau)=0$, where
$x(t)|_{t\ra t_c^*-0}=+\infty$. These points are:
$t_c^*=11.4328$ (for solid line), $t_c^*=2.87170$ (for
dashed line), $t_c^*=3.33026$ (for dotted line), and
$t_c^*=3.83074$ (for dashed-dotted line).

\vskip 5mm

{\bf Fig. 22}. Solutions $x(t)$ to Eq. (\ref{eq61}) as
functions of time for the parameters $\tau=0.48$ (solid
line) and $\tau=0.485$ (dashed-dotted line), where
$\tau<\tau_0=0.495125$. Other parameters are: $a=2$,
$x_0=1$, and $b=-2.5$. The solutions $x(t)$ tend
non-monotonically to their stationary point
$x_2^*=-a/(b+1)=4/3$ as $t\ra\infty$.

\vskip 5mm

{\bf Fig. 23}. Solutions $x(t)$ to Eq. (\ref{eq81})
(solid line) and to the corresponding equation obtained
by linearizing Eq. (\ref{eq81}) around the fixed point
$x^*=0$ (dashed-dotted line). Parameters for the
equations are: $a=0$, $b=2$, and $\tau=4$ (note that
$\tau>\tau_0=\pi$). The figure shows that the solution
to the linearized equation is unstable for $\tau>\tau_0$,
as the stability analysis prescribes, whereas the solution
to the nonlinear equation is stable for $\tau>\tau_0$,
tending to its stationary point $x^*=0$ as $t\ra\infty$.

\vskip 5mm

{\bf Fig. 24}. Solutions $x(t)$ to Eq. (\ref{eq81}) as
functions of time for the parameters $b=0.55$ (solid
line) and $b=5$ (dashed-dotted line). Other parameters
are: $a=1$, $x_0=2$, and $\tau=20$. The solutions $x(t)$
decrease monotonically to their stationary point $x^*=0$
as $t\ra\infty$.

\vskip 5mm

{\bf Fig. 25}. Solutions $x(t)$ to Eq. (\ref{eq81}) as
functions of time for the parameters $b=-5$ (solid line),
$b=-7.5$ (dashed line), and $b=-10$ (dashed-dotted line).
Other parameters are: $a=2$, $x_0=1$, and $\tau=0.5$.
The solutions $x(t)$ decrease monotonically with a sharp
but continuous drop ending at $0$ at time $t_d=1.24290$
(for solid line), $t_d=1.66635$ (for dashed line), and
$t_d=1.97114$ (for dashed-dotted line) defined by the
equation $a+bx(t_d-\tau)=0$. At these moments, of time
$\dot{x}(t)|_{t\ra t_d-0}=-\infty$.

\vskip 5mm

{\bf Fig. 26}. Scheme of the variety of qualitatively
different solution types for the most complicated and the
most realistic regime of Sec. 4, when gain (birth) prevails 
over loss (death) and competition is stronger than 
cooperation.

\vskip 5mm

{\bf Fig. 27}. Scheme of qualitatively different solution 
types for the case of Sec. 5, when gain prevails over 
loss and cooperation prevails over competition.

\vskip 5mm

{\bf Fig. 28}. Summarizing scheme of qualitatively different 
solution types for the case of Sec. 6, when loss prevails over 
gain and competition prevails over cooperation.

\vskip 5mm

{\bf Fig. 29}. Summarizing scheme of qualitatively different 
solution types for the case of Sec. 7, when loss prevails over 
gain and cooperation prevails over competition.



\newpage

\begin{figure}[h]
\centerline{\includegraphics[width=16cm]{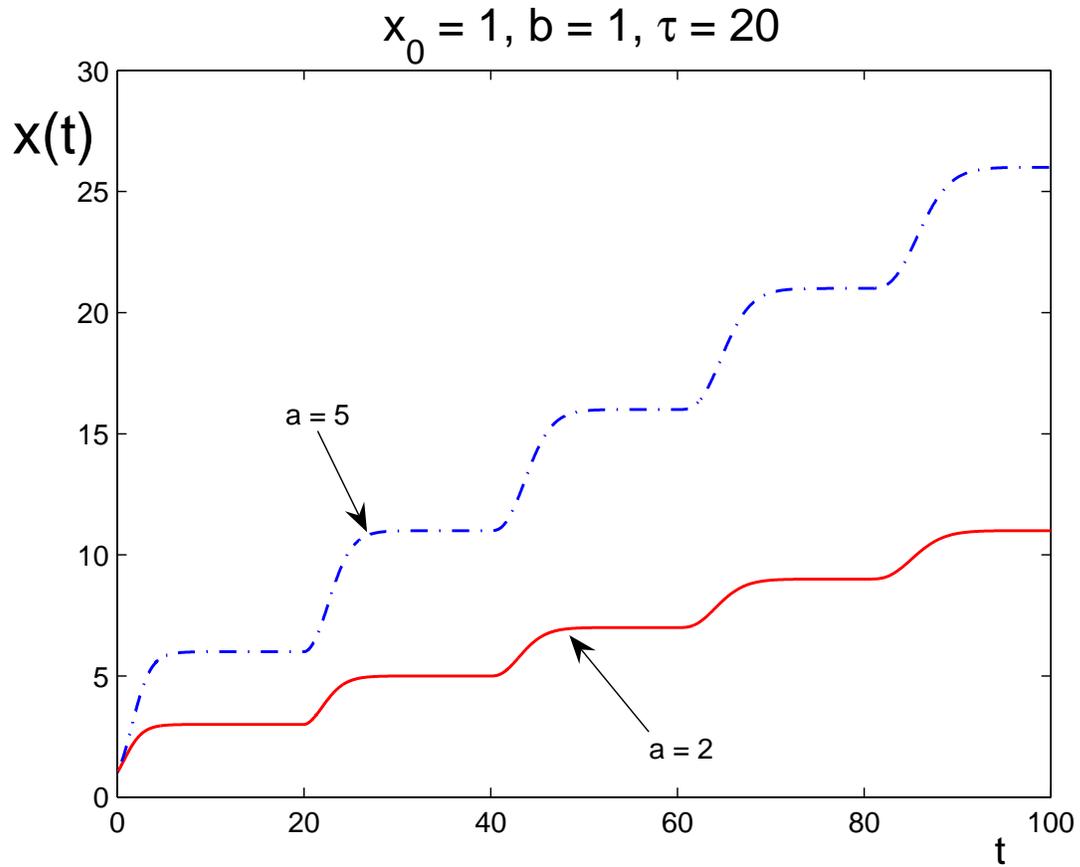}}
\caption{Solutions $x(t)$ to Eq. (\ref{eq31}) as
functions of time for the parameters $a=2$ (solid line)
and $a=5$ (dashed-dotted line). Other parameters are:
$x_0=1$, $b=1$, and $\tau=20$. Solutions grow monotonically
by steps of duration $\simeq \tau$, and $x(t)\ra+\infty$ as
$t\ra+\infty$.}
\label{fig:Fig.1}
\end{figure}

\newpage

\begin{figure}[h]
\centerline{\includegraphics[width=16cm]{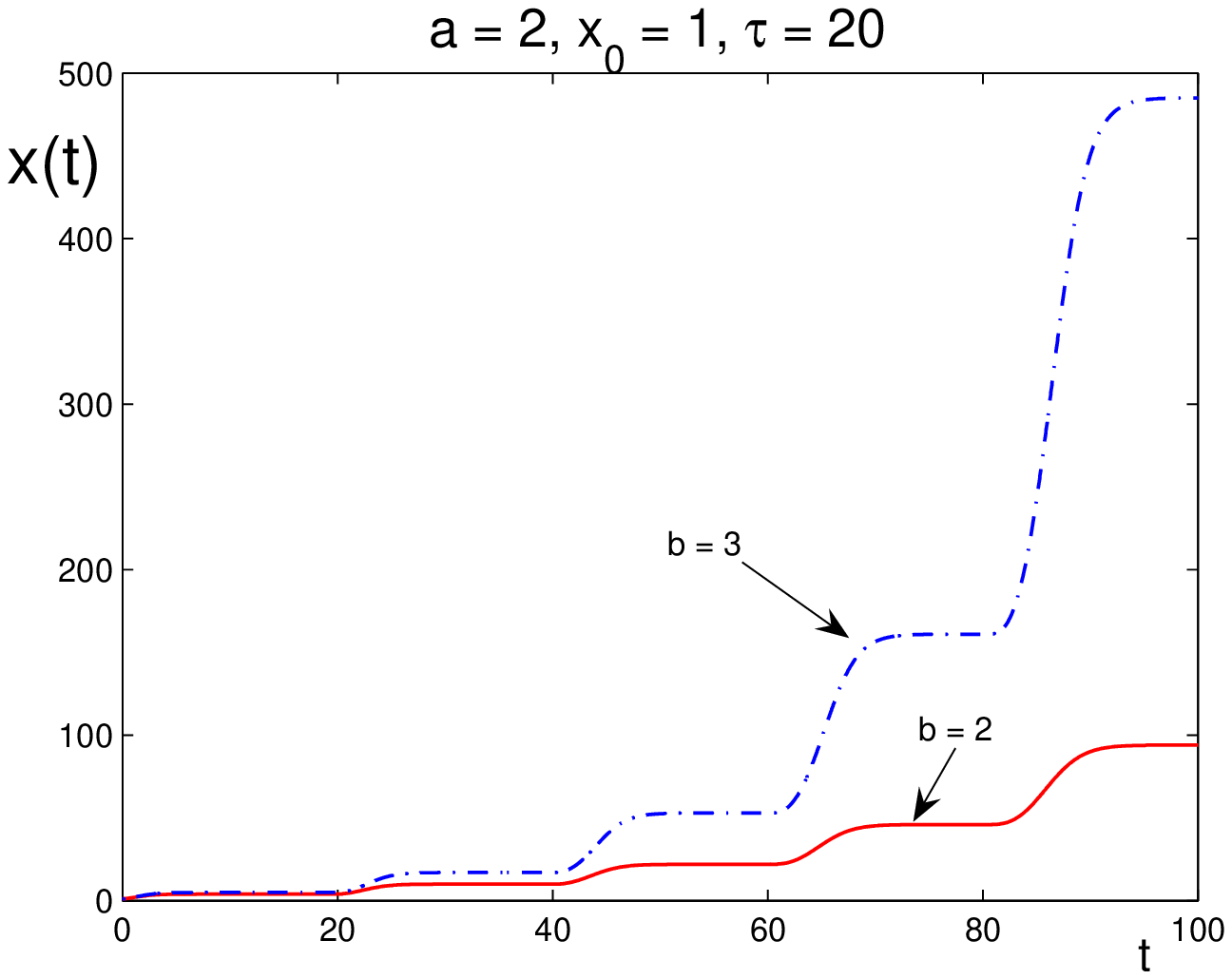}}
\caption{Solutions $x(t)$ to Eq. (\ref{eq31}) as
functions of time for the parameters $b=2$ (solid line)
and $b=3$ (dashed-dotted line). Other parameters are:
$a=2$, $x_0=1$, and $\tau=20$. Solutions grow monotonically
by steps  of duration $\simeq \tau$, and $x(t)\ra+\infty$ as
$t\ra\infty$.}
\label{fig:Fig.2}
\end{figure}

\newpage

\begin{figure}[h]
\centerline{\includegraphics[width=16cm]{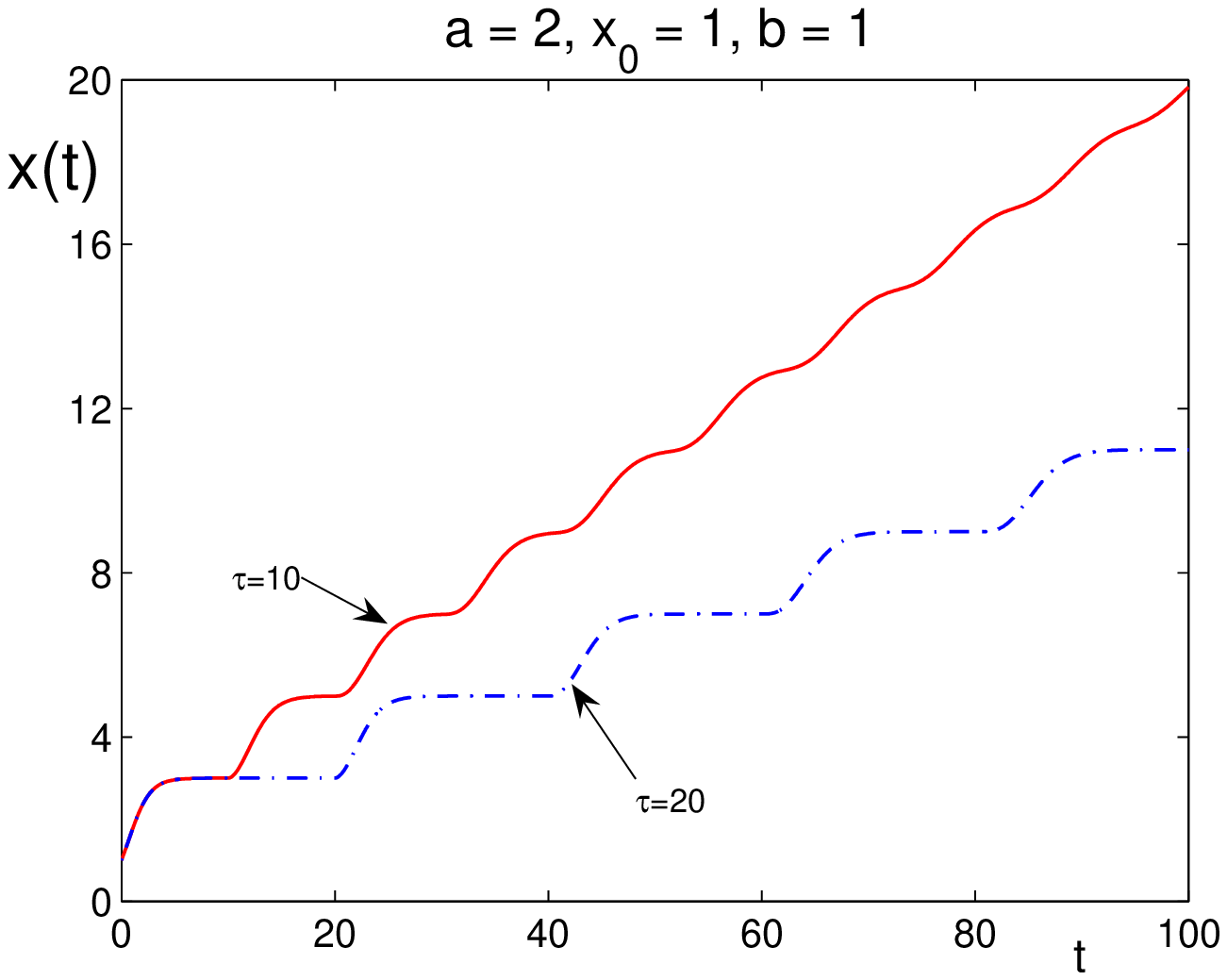}}
\caption{Solutions $x(t)$ to Eq. (\ref{eq31}) as
functions of time for the parameters $\tau=10$ (solid line)
and $\tau=20$ (dashed-dotted line). Other parameters are:
$a=2$, $x_0=1$, and $b=1$. Solutions grow monotonically by
steps  of duration $\simeq \tau$, and $x(t)\ra+\infty$ as
$t\ra\infty$.}
\label{fig:Fig.3}
\end{figure}

\newpage

\begin{figure}[h]
\centerline{\includegraphics[width=16cm]{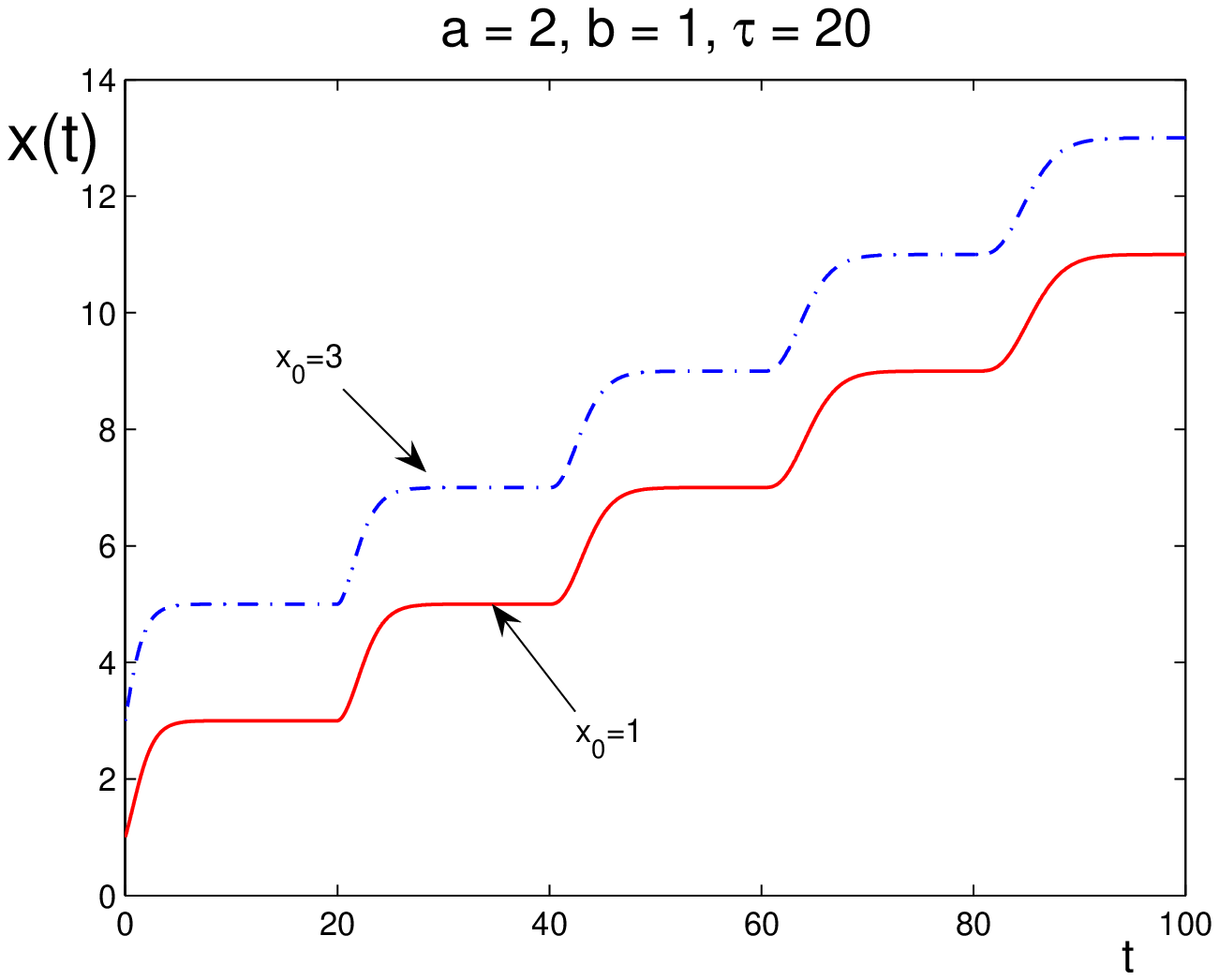}}
\caption{Solutions $x(t)$ to Eq. (\ref{eq31}) as
functions of time for the parameters $x_0=1$ (solid line)
and $x_0=3$ (dashed-dotted line). Other parameters are:
$a=2$, $b=1$, and $\tau=20$. Solutions grow monotonically
by steps  of duration $\simeq \tau$, and $x(t)\ra+\infty$ as
$t\ra\infty$.}
\label{fig:Fig.4}
\end{figure}

\newpage

\begin{figure}[h]
\centerline{\includegraphics[width=16cm]{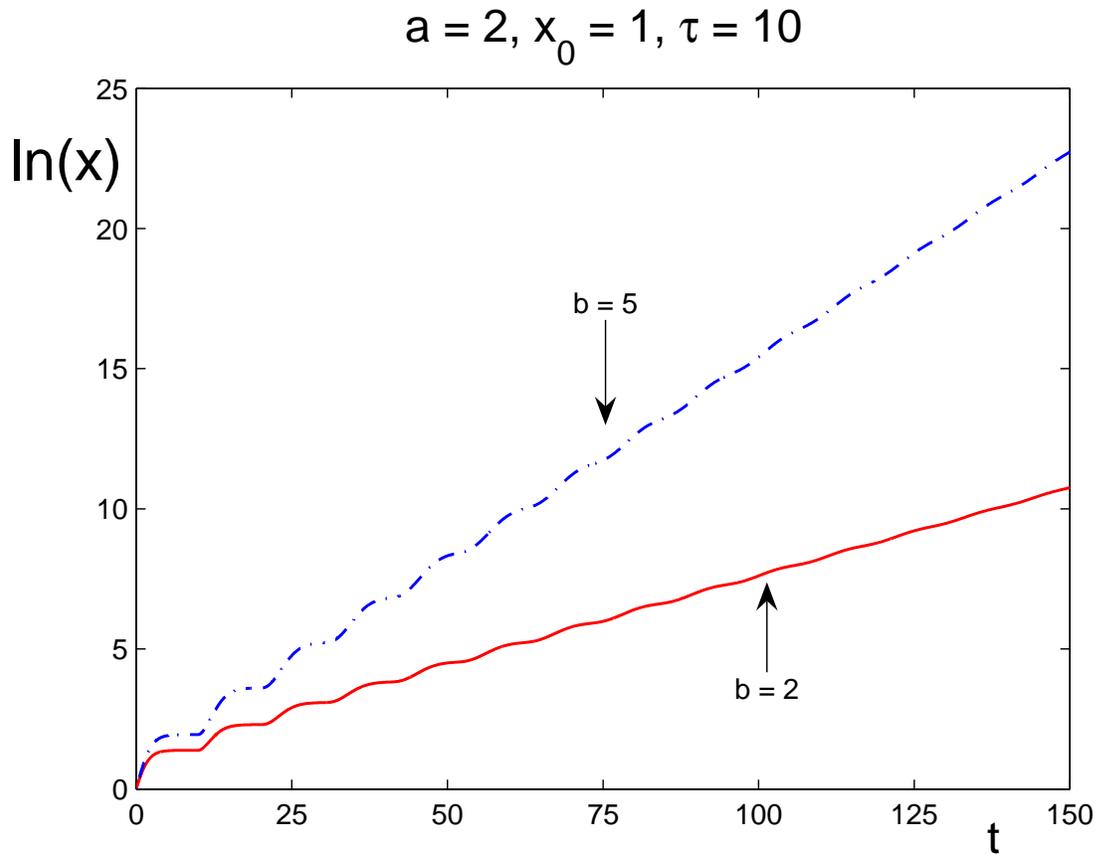}}
\caption{Logarithm of the solutions $x(t)$
to Eq. (\ref{eq31}) as functions of time for the parameters
$b=2$ (solid line) and $b=5$ (dashed-dotted line)
exemplifying their average long-term exponential growth. Other
parameters are: $a=2$, $x_0=1$, and $\tau=10$.}
\label{fig:Fig.5}
\end{figure}

\newpage

\begin{figure}[h]
\centerline{\includegraphics[width=16cm]{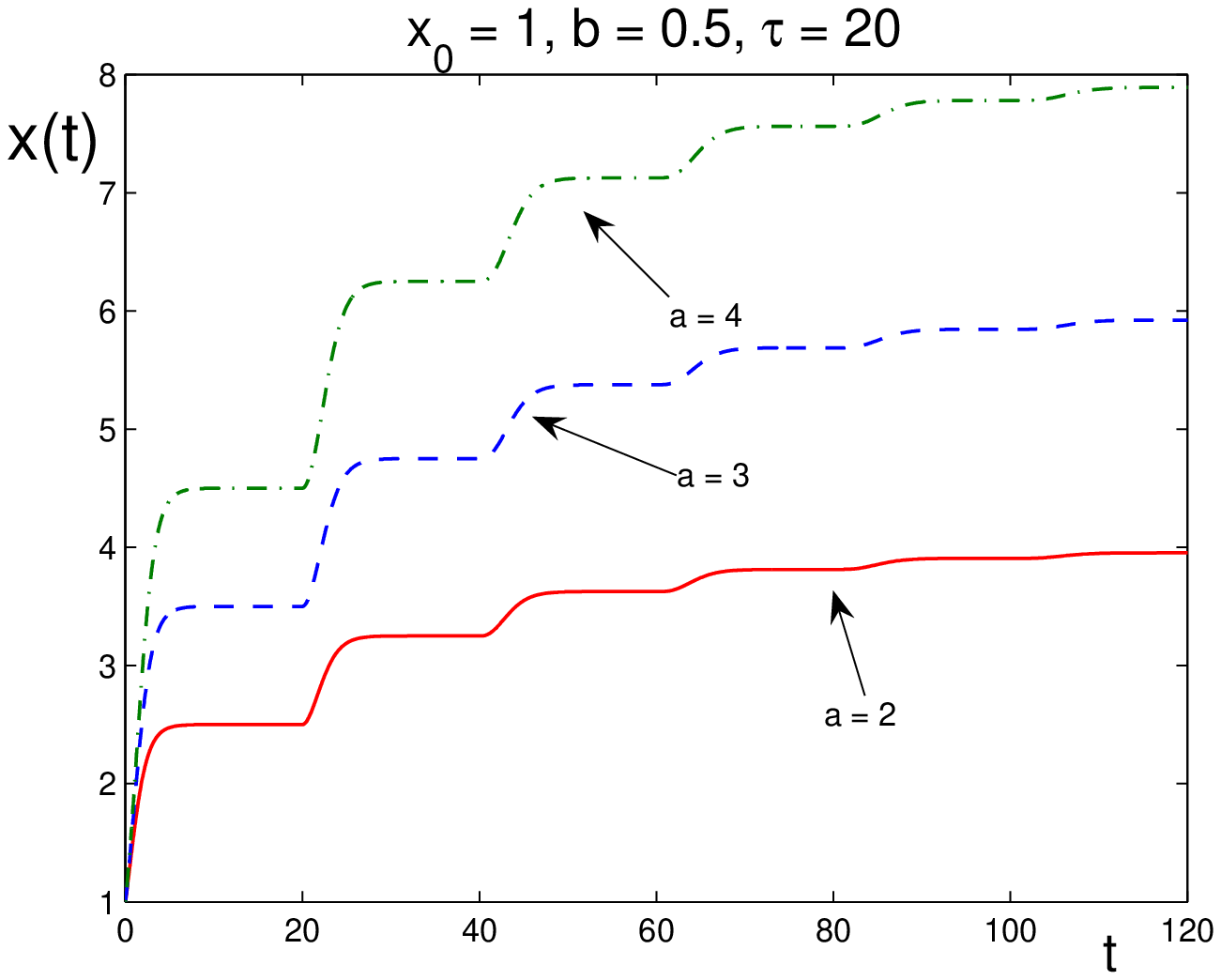}}
\caption{Solutions $x(t)$ to Eq. (\ref{eq31}) as
functions of time for the parameters $a=2$ (solid line),
$a=3$ (dashed line), and $a=4$ (dashed-dotted line).
Other parameters are: $x_0=1$, $b=0.5$, and $\tau=20$.
The solutions $x(t)$ monotonically grow by steps to their
stationary points $x_2^*=a/(1-b)$ as $t\ra\infty$.
Stationary points are: $x_2^*=4$ (for solid line), $x_2^*=6$
(for dashed line), and $x_2^*=8$ (for dashed-dotted line).}
\label{fig:Fig.6}
\end{figure}

\newpage

\begin{figure}[h]
\centerline{\includegraphics[width=16cm]{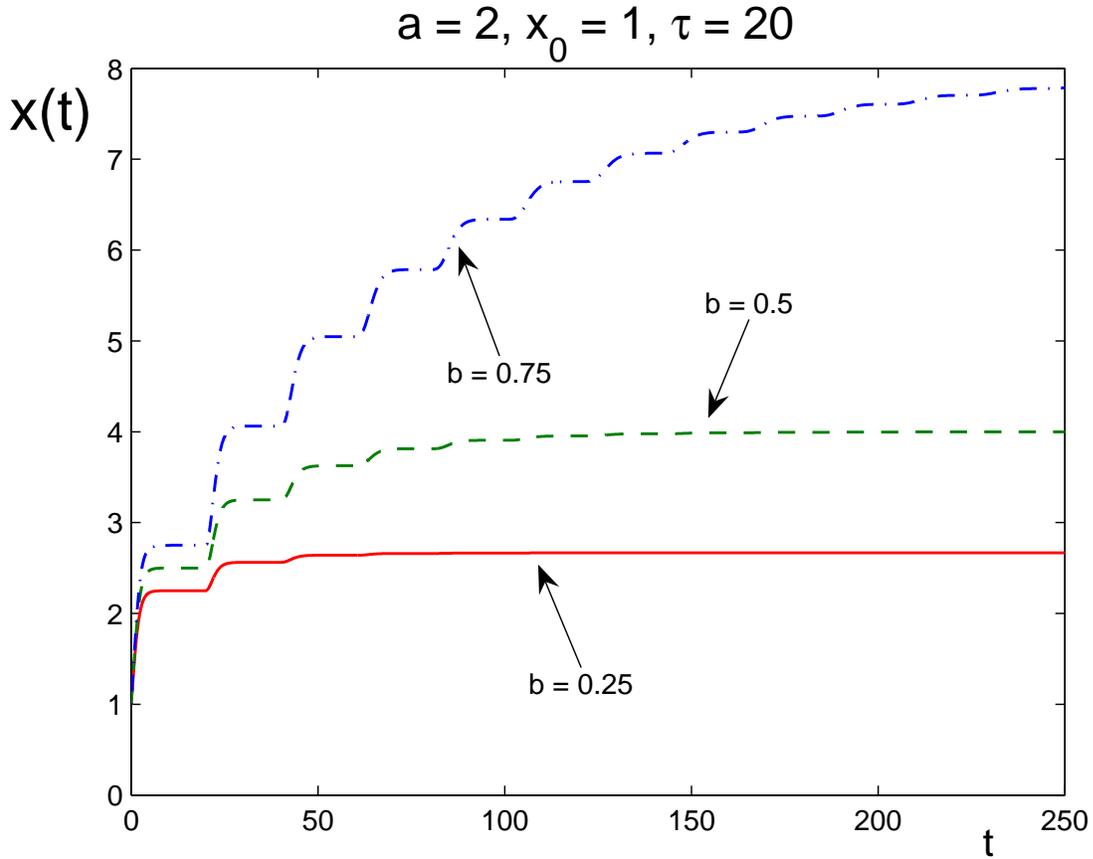}}
\caption{Solutions $x(t)$ to Eq. (\ref{eq31}) as
functions of time for the parameters $b=0.25$ (solid line),
$b=0.5$ (dashed line), and $b=0.75$ (dashed-dotted line).
Other parameters are: $a=2$, $x_0=1$, and $\tau=20$.
The solutions $x(t)$ monotonically grow by steps to their
stationary points $x_2^*=a/(1-b)$ as $t\ra\infty$.
Stationary points are: $x_2^*=8/3\approx 2.67$ (for
solid line), $x_2^*=4$ (for dashed line), and $x_2^*=8$
(for dashed-dotted line).}
\label{fig:Fig.7}
\end{figure}

\newpage

\begin{figure}[h]
\centerline{\includegraphics[width=16cm]{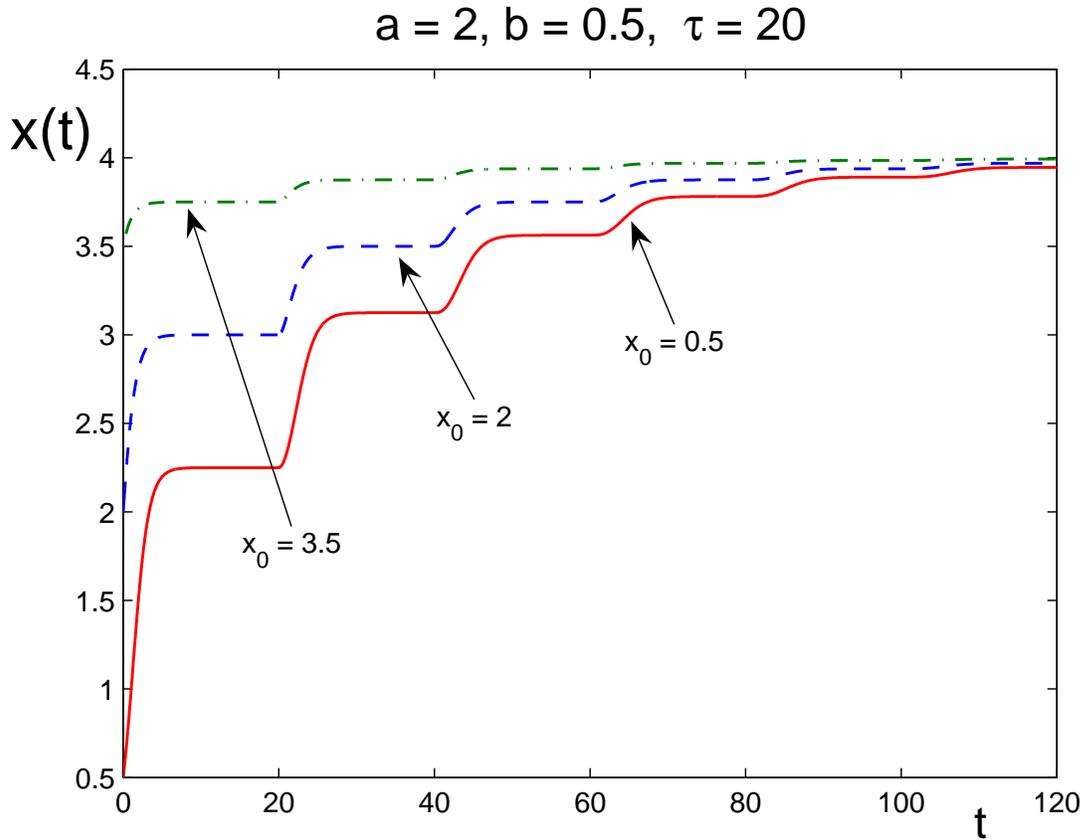}}
\caption{Solutions $x(t)$ to Eq. (\ref{eq31}) as
functions of time for the parameters $x_0=0.5$ (solid
line), $x_0=2$ (dashed line), and $x_0=3.5$ (dashed-dotted
line). Other parameters are: $a=2$, $b=0.5$, and $\tau=20$.
The solutions $x(t)$ monotonically grow by steps to their
stationary point $x_2^*=a/(1-b)=4$ as $t\ra\infty$.}
\label{fig:Fig.8}
\end{figure}
\newpage

\begin{figure}[h]
\centerline{\includegraphics[width=16cm]{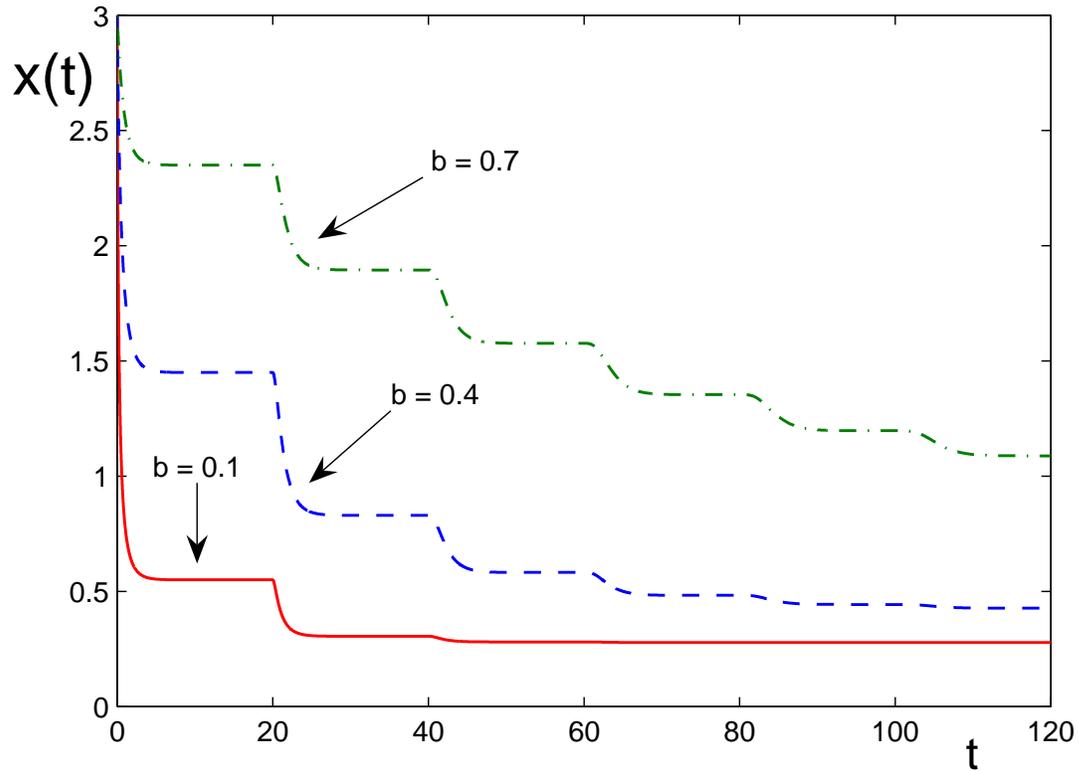}}
\caption{Solutions $x(t)$ to Eq. (\ref{eq31}) as
functions of time for the parameters $b=0.1$ (solid line),
$b=0.4$ (dashed line), and $b=0.7$ (dashed-dotted line).
Other parameters are: $a=0.25$, $x_0=3$, and $\tau=20$.
The solutions $x(t)$ monotonically decrease by steps to
their stationary points $x_2^*=a/(1-b)$ as $t\ra\infty$.
Stationary points are: $x_2^*=5/18\approx 0.278$ (for solid
line), $x_2^*=5/12\approx 0.417$ (for dashed line), and
$x_2^*=5/6\approx 0.833$ (for dashed-dotted line).}
\label{fig:Fig.9}
\end{figure}

\newpage

\begin{figure}[h]
\centerline{\includegraphics[width=16cm]{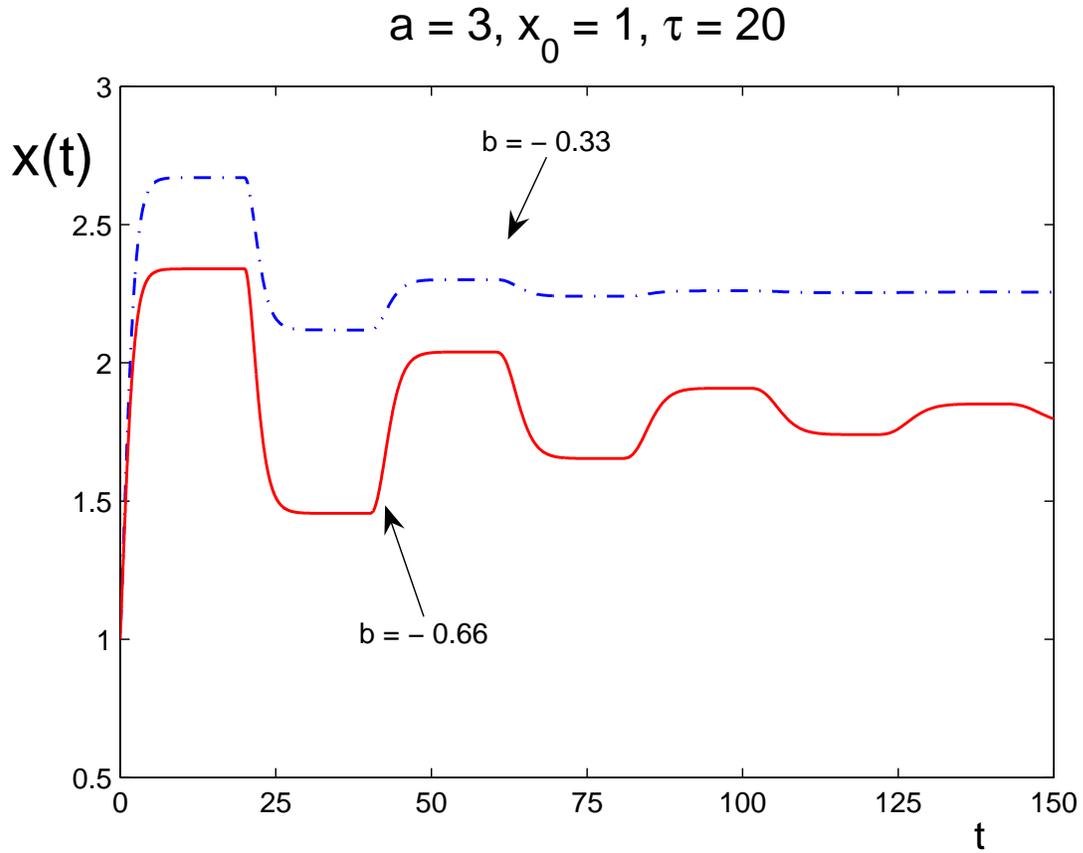}}
\caption{Solutions $x(t)$ to Eq. (\ref{eq31}) as
functions of time for the parameters $b=-0.66$ (solid line)
and $b=-0.33$ (dashed-dotted line). Other parameters are:
$a=3$, $x_0=1$, and $\tau=20$. The solutions $x(t)$
non-monotonically grow by steps to their stationary points
$x_2^*=a/(1-b)$ as $t\ra\infty$. The stationary points are:
$x_2^*=1.80723$ (for solid line) and $x_2^*=2.25564$
(for dashed-dotted line).}
\label{fig:Fig.10}
\end{figure}

\newpage

\begin{figure}[h]
\centerline{\includegraphics[width=16cm]{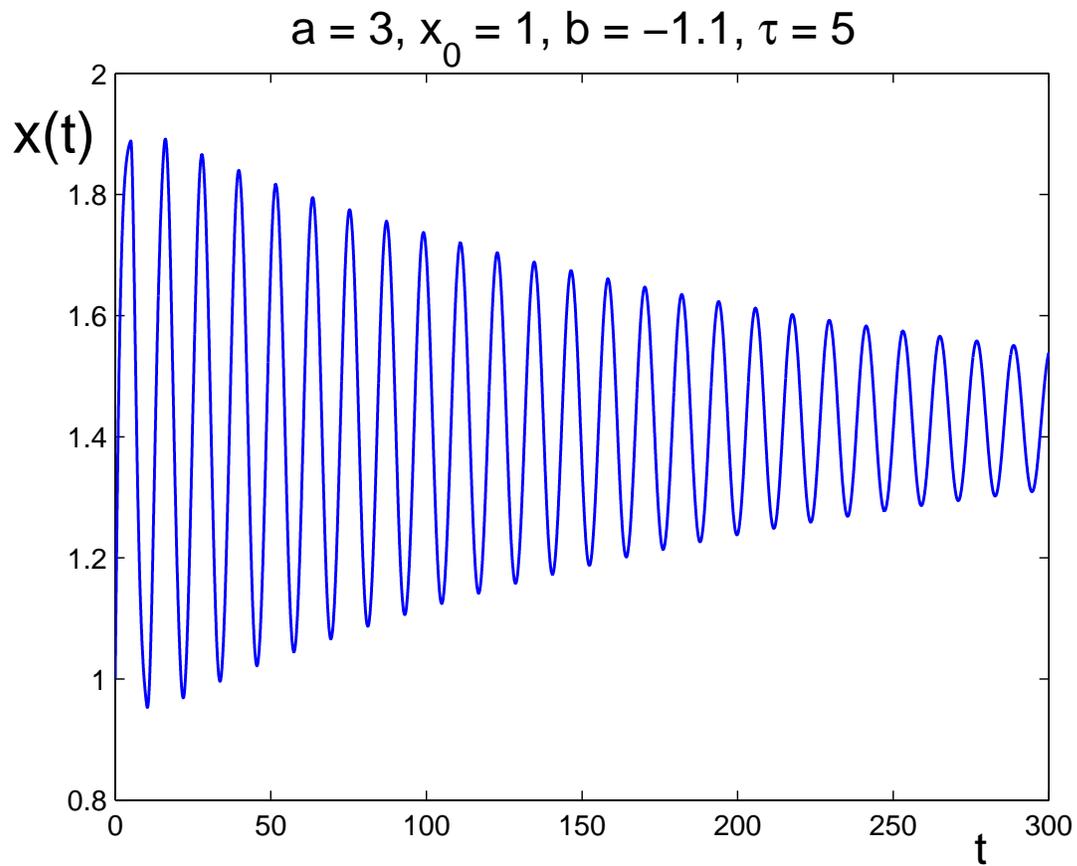}}
\caption{Solution $x(t)$ to Eq. (\ref{eq31}) as a
function of time for the parameter $\tau=5<\tau_0=5.91784$.
Other parameters are: $a=3$, $x_0=1$, and $b=-1.1$. The solution
$x(t)$ tends in an oscillatory manner to its stationary point
$x_2^*=a/(1-b)=1.42857$ as $t\ra\infty$.}
\label{fig:Fig.11}
\end{figure}

\newpage

\begin{figure}[h]
\centerline{\includegraphics[width=16cm]{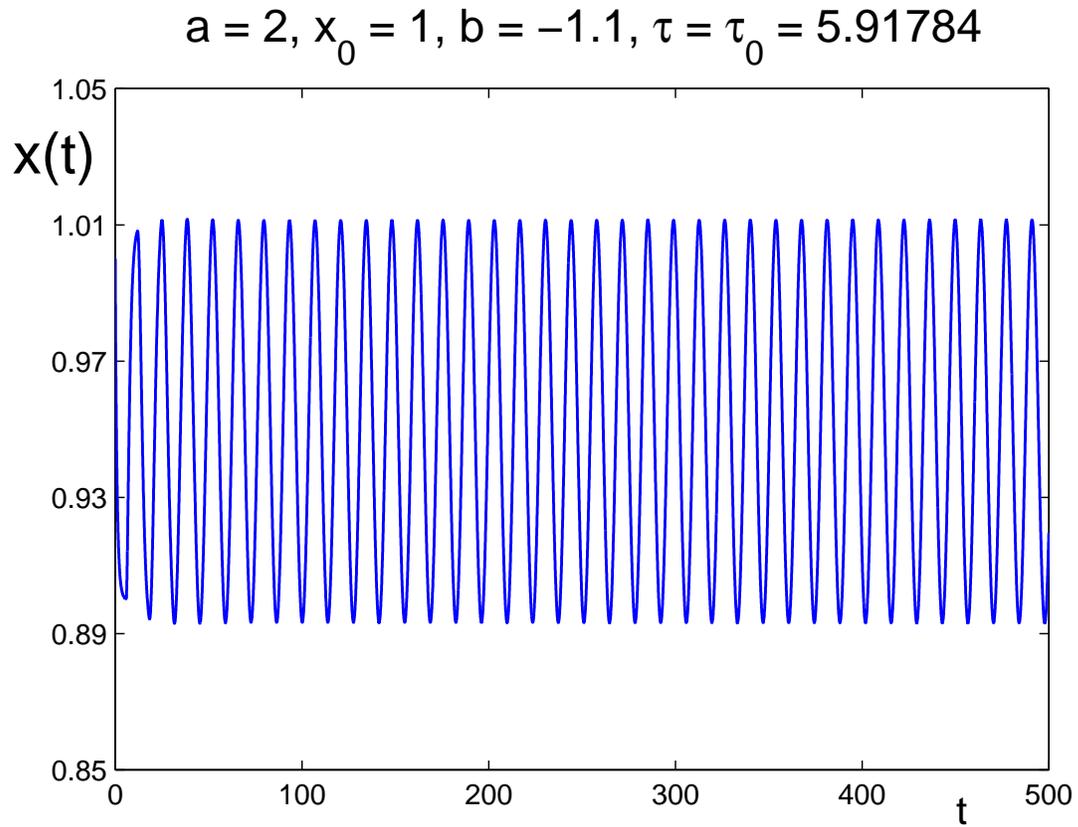}}
\caption{Solution $x(t)$ to Eq. (\ref{eq31}) as a
function of time for the parameter $\tau=\tau_0=5.91784$.
Other parameters are: $a=2$, $x_0=1$, and $b=-1.1$. The solution
$x(t)$ oscillates around its stationary point
$x_2^*=a/(1-b)=0.952381$ as $t\ra\infty$.}
\label{fig:Fig.12}
\end{figure}

\newpage

\begin{figure}[h]
\centerline{\includegraphics[width=16cm]{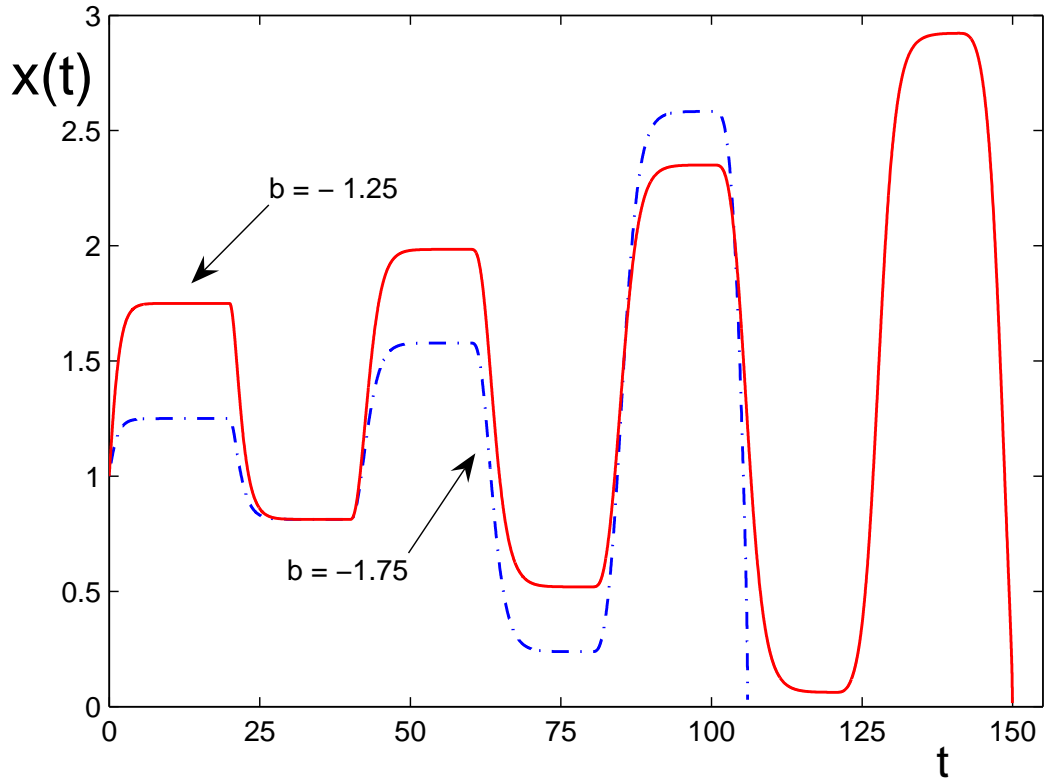}}
\caption{Solutions $x(t)$ to Eq. (\ref{eq31}) as
functions of time for the parameters $b=-1.25$ (solid line)
and $b=-1.75$ (dashed-dotted line). Other parameters are:
$a=3$, $x_0=1$, and $\tau=20$. The solutions $x(t)$ oscillate
with increasing amplitude until time
$t_d=149.932$ (for solid line) and $t_d=105.972$ (for
dashed-dotted line) defined by the equation
$a+bx(t_d-\tau)=0$, at which
$\dot{x}(t)|_{t\ra t_d-0}=-\infty$.}
\label{fig:Fig.13}
\end{figure}

\newpage

\begin{figure}[h]
\centerline{\includegraphics[width=16cm]{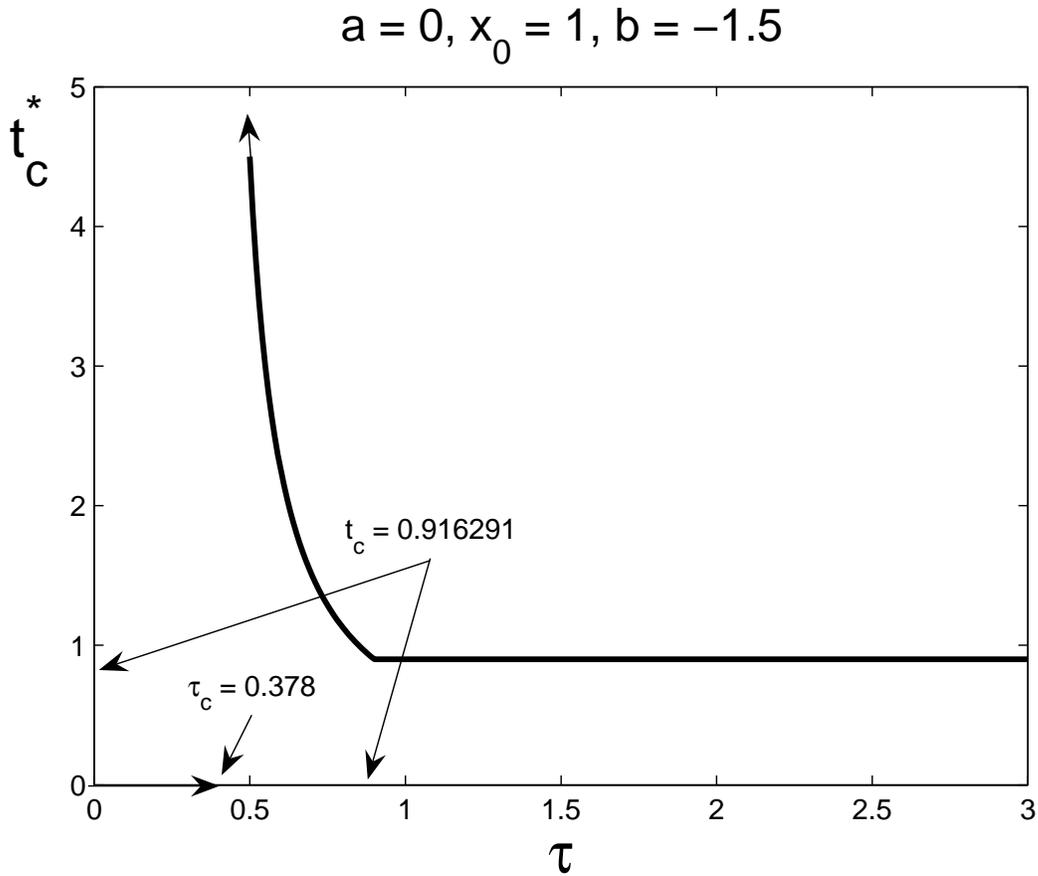}}
\caption{Dependence of the critical time $t_c^*(\tau)$,
where the function $x(t)$ exhibits a singularity, as a
function of the lag $\tau$. Here the parameters are: $a=0$,
$x_0=1$, and $b=-1.5$.  At time $t=t_c^*$, the solution has
a singularity: $x(t)|_{t\ra t_c^*-0}=+\infty$. The instant
$t_c^*$ is defined by $a+bx(t_c^*-\tau)=0$. The time
$t_c=\ln(1-y_0/x_0)=0.916291$ is the point of singularity.}
\label{fig:Fig.14}
\end{figure}

\newpage

\begin{figure}[h]
\centerline{\includegraphics[width=16cm]{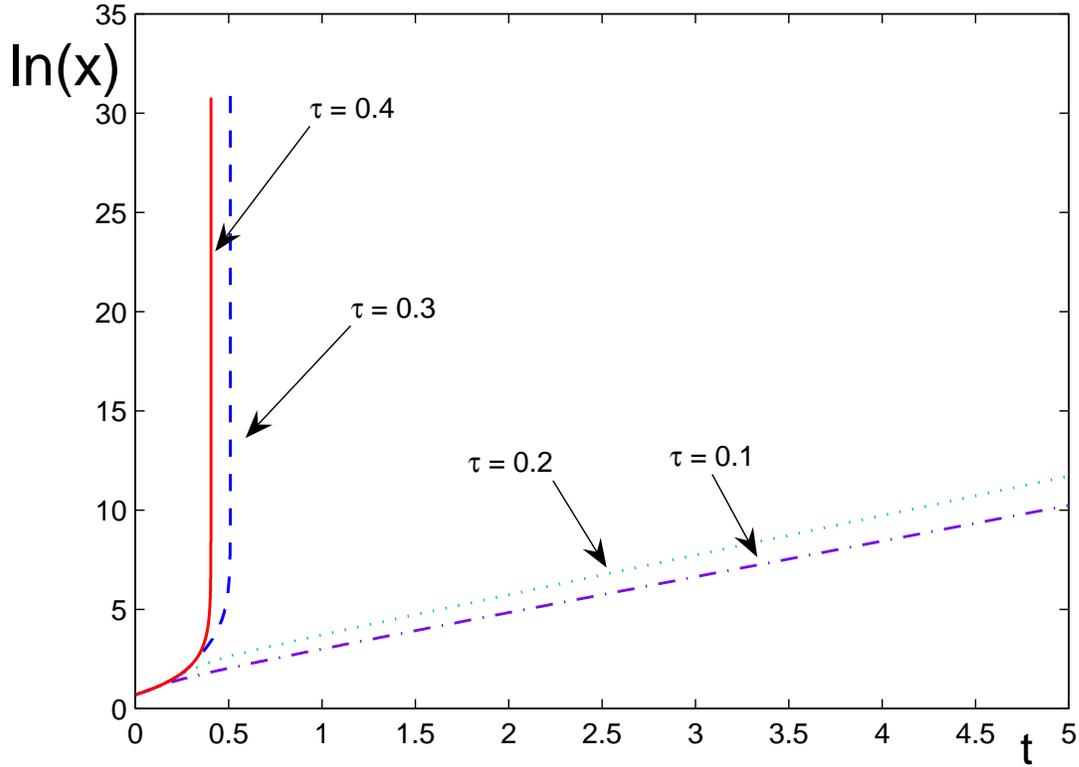}}
\caption{Solutions $x(t)$ to Eq. (\ref{eq31}) as
functions of time for the parameters $\tau=0.1$ (dashed-dotted
line), $\tau=0.2$ (dotted line), $\tau=0.3$ (dashed line), and
$\tau=0.4$ (solid line). The other parameters are: $a=2$, $x_0=2$,
and $b=-1.5$. For all $\tau>t_c=\ln(1-y_0/x_0)=0.405465$, we have
$t_c^*=t_c$. If $\tau\leq t_c$, there exists $\tau_c\approx
0.27217$ such that if $0<\tau\leq\tau_c$, then $x(t)$ grows
exponentially to $+\infty$. If $\tau_c<\tau\leq t_c$, then
there exists a point of singularity $t_c^*>t_c$, defined by
$a+bx(t_c^*-\tau)=0$, such that $x(t)|_{t\ra t_c^*-0}=+\infty$.
When $\tau\to t_c-0$, then $t_c^*\ra t_c+0$. The values of $t_c^*$
are respectively $t_c^*=0.40560$ (for solid line, $\tau=0.4$) and
$t_c^*=0.50993$ (for dashed line, $\tau=0.3$).}
\label{fig:Fig.15}
\end{figure}

\newpage

\begin{figure}[h]
\centerline{\includegraphics[width=16cm]{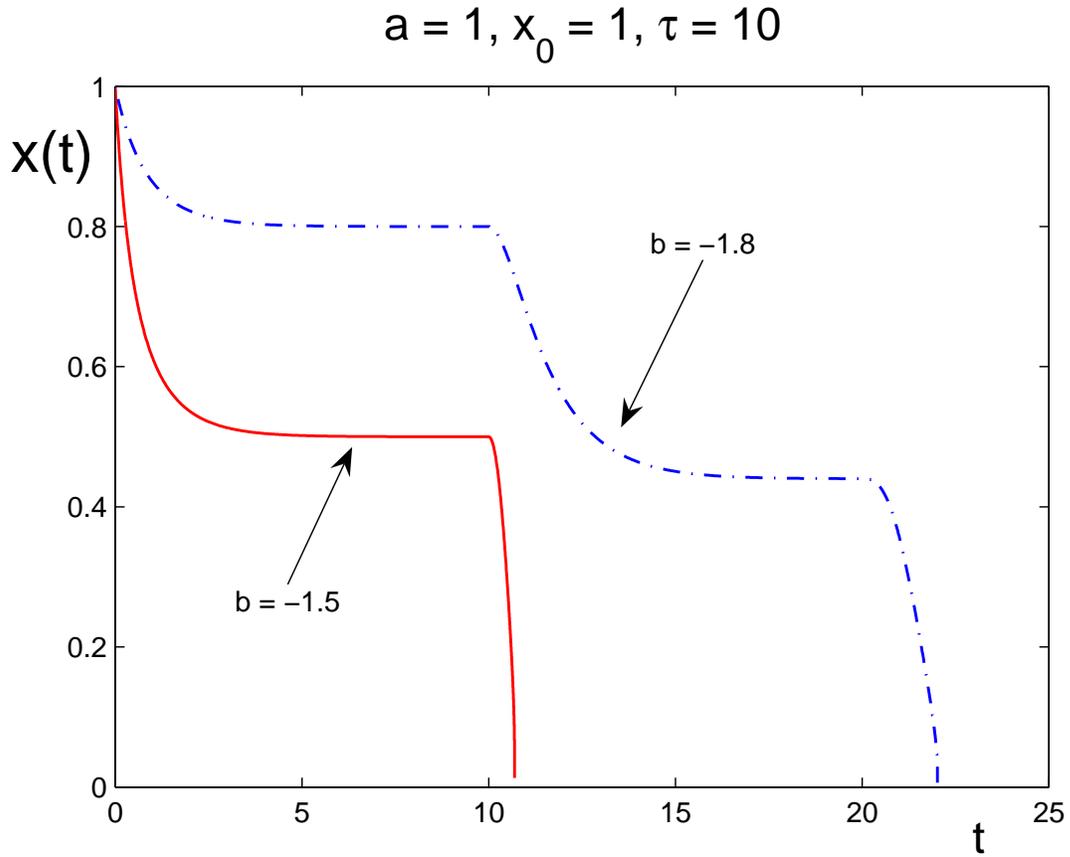}}
\caption{Solutions $x(t)$ to Eq. (\ref{eq51}) as
functions of time for the parameters $b=-1.5$ (solid line)
and $b=-1.8$ (dashed-dotted line). Other parameters are:
$a=1$, $x_0=1$, and $\tau=10$. The solutions $x(t)$ decrease
by steps until time $t_d=10.6932$ (for solid
line) and $t_d=22.0171$ (for dashed-dotted line) defined by
the equation $a+bx(t_d-\tau)=0$. At these times,
$\dot{x}(t)|_{t\ra t_d-0}=-\infty$.}
\label{fig:Fig.16}
\end{figure}

\newpage

\begin{figure}[h]
\centerline{\includegraphics[width=16cm]{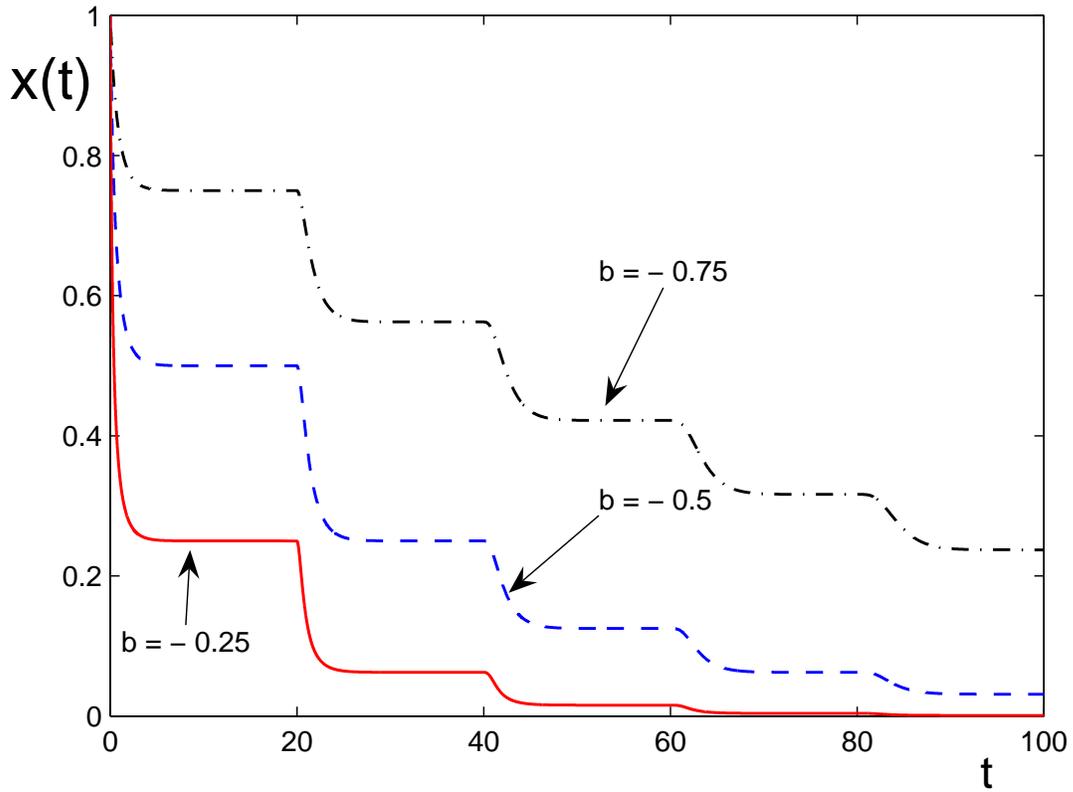}}
\caption{Solutions $x(t)$ to Eq. (\ref{eq51}) as
functions of time for the parameters $b=-0.25$ (solid line),
$b=-0.5$ (dashed line), and $b=-0.75$ (dashed-dotted line).
Other parameters are: $a=0$, $x_0=1$, and $\tau=20$.
The solutions $x(t)$ monotonically decrease by steps to their
stationary point $x^*=0$ as $t\ra\infty$.}
\label{fig:Fig.17}
\end{figure}

\newpage

\begin{figure}[h]
\centerline{\includegraphics[width=16cm]{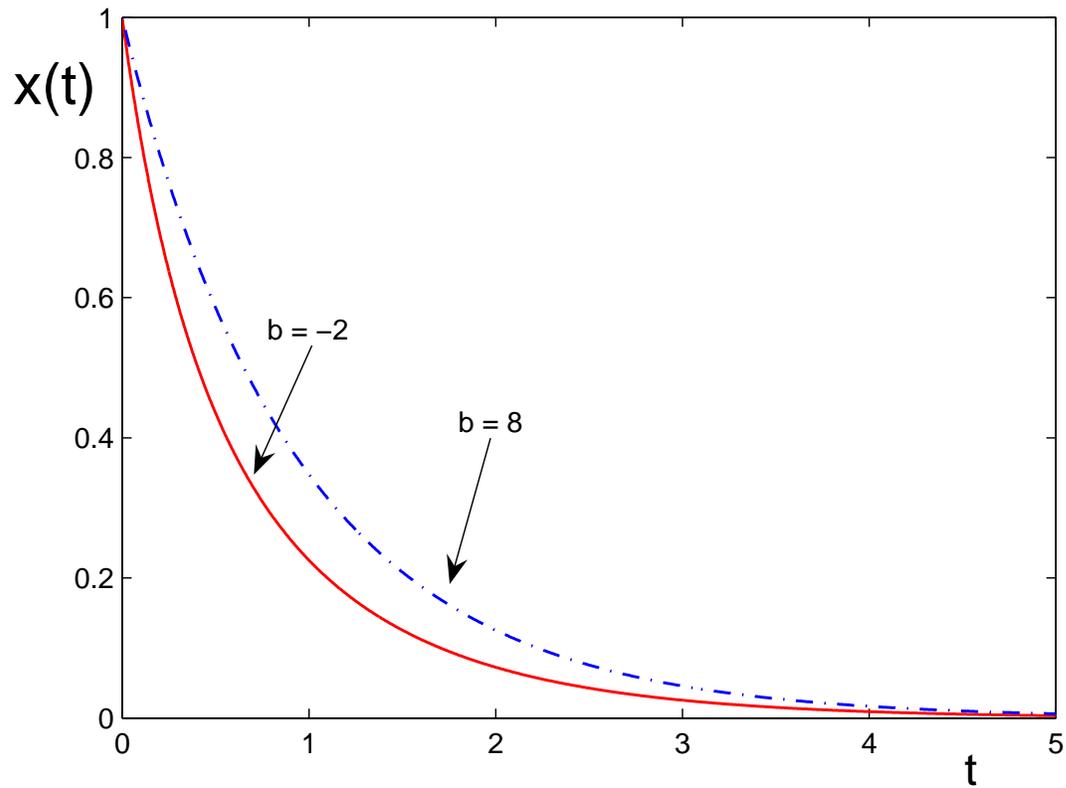}}
\caption{Solutions $x(t)$ to Eq. (\ref{eq61}) as
functions of time for the parameters $b=-2$ (solid line)
and $b=8$ (dashed-dotted line). Other parameters are:
$a=3$, $x_0=1$, and $\tau=10$. The solutions $x(t)$
monotonically decrease to their stationary point
$x^*=0$ as $t\ra\infty$.}
\label{fig:Fig.18}
\end{figure}

\newpage

\begin{figure}[h]
\centerline{\includegraphics[width=16cm]{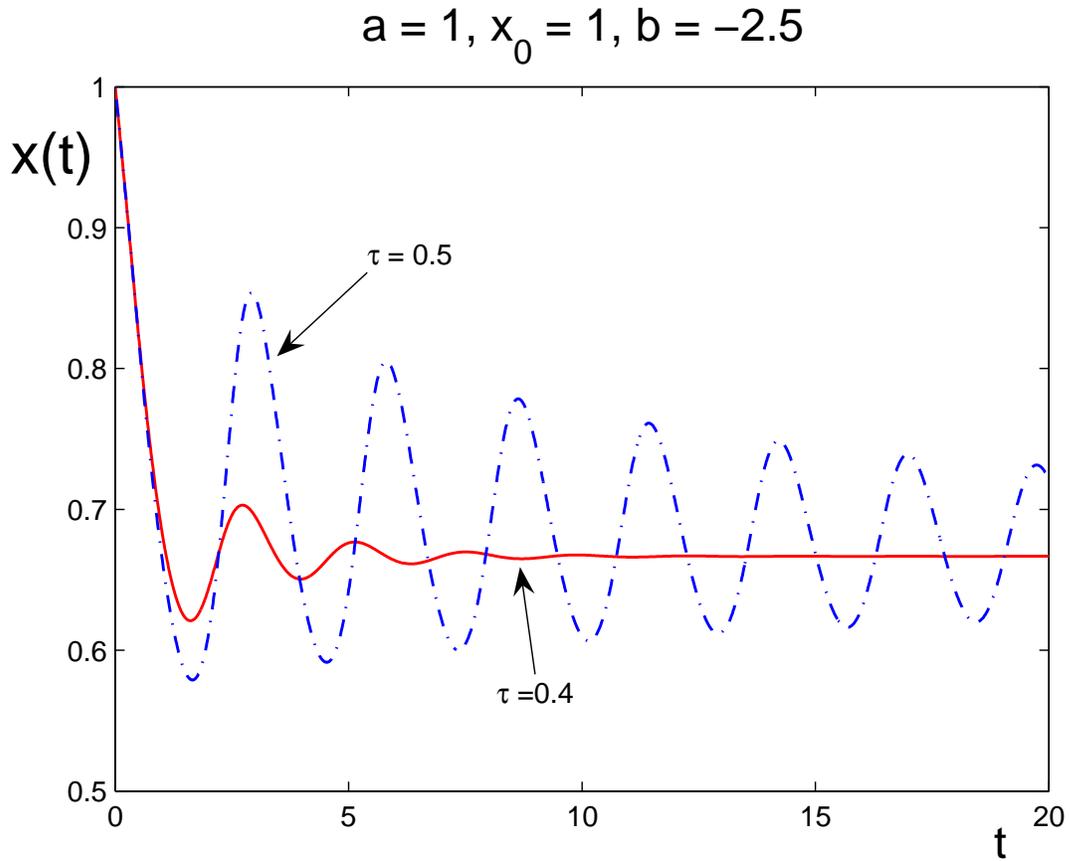}}
\caption{Solutions $x(t)$ to Eq. (\ref{eq61}) as
functions of time for the parameters $\tau=0.4$ (solid
line) and $\tau=0.5$ (dashed-dotted line), where
$\tau<\tau_0=0.505951$. Other parameters are: $a=1$,
$x_0=1$, and $b=-2.5$. The solutions $x(t)$ converge by
oscillating to their stationary point $x_2^*=-a/(b+1)=2/3$
as $t\ra\infty$.}
\label{fig:Fig.19}
\end{figure}

\newpage

\begin{figure}[h]
\centerline{\includegraphics[width=16cm]{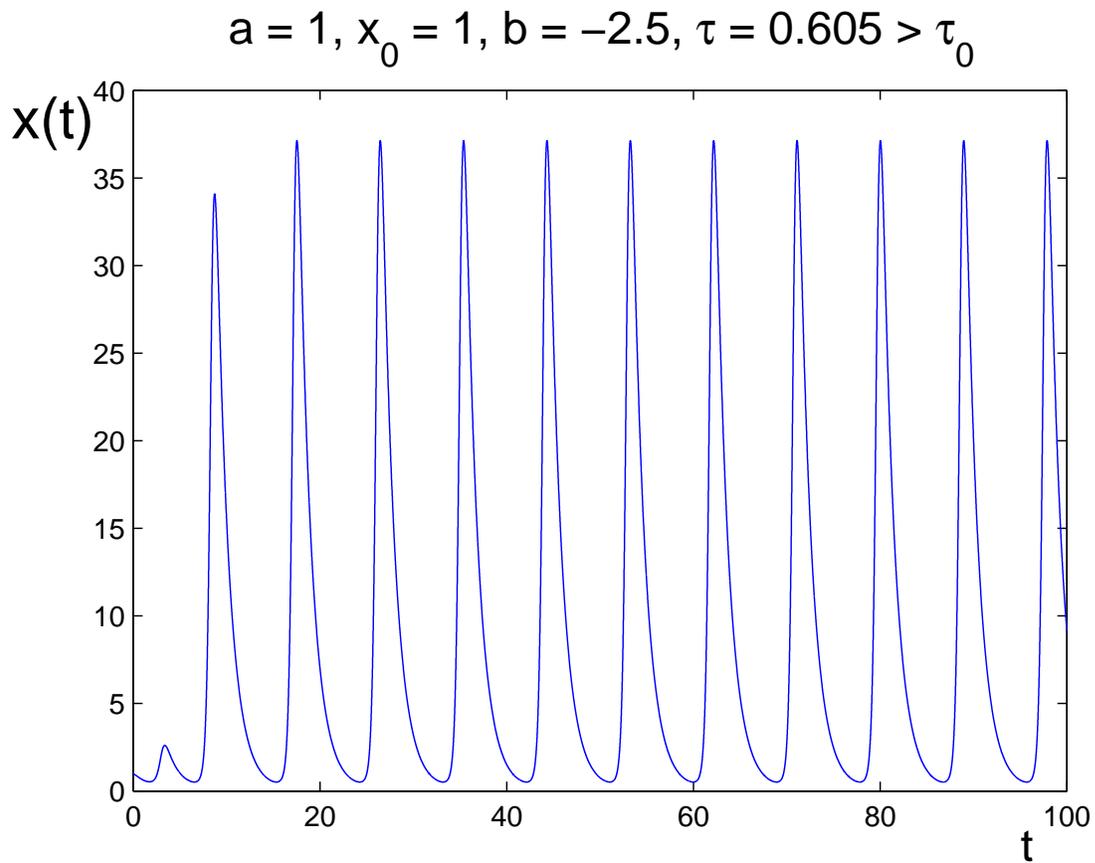}}
\caption{Solution $x(t)$ to Eq. (\ref{eq61}) as
a function of time for the parameter $\tau=0.605>\tau_0$,
where $\tau_0=0.505951$. Other parameters are the same as
for Fig. 19. The solution $x(t)$ exhibits sustained oscillations
with an amplitude which is an increasing function of the delay
time $\tau$ and a period much larger than $\tau$.}
\label{fig:Fig.20}
\end{figure}

\newpage

\begin{figure}[h]
\centerline{\includegraphics[width=16cm]{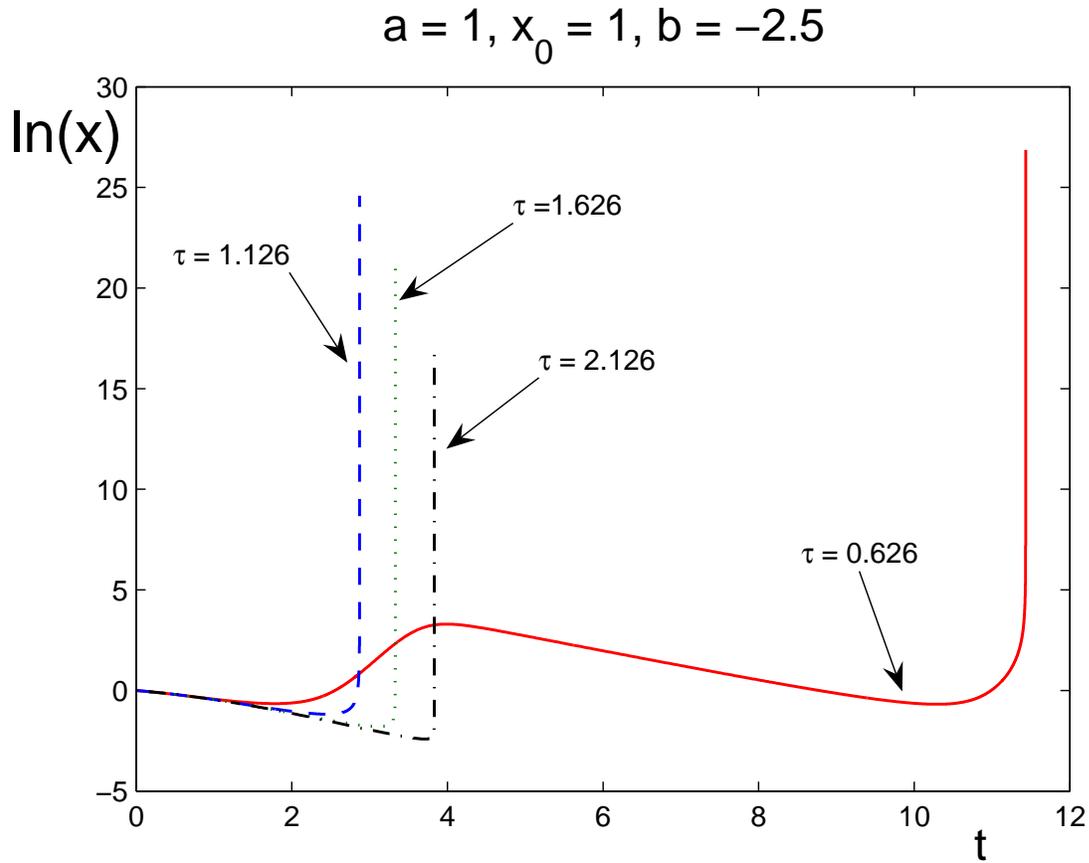}}
\caption{Logarithmic behavior of solutions $x(t)$
to Eq. (\ref{eq61}) as functions of time for the parameters
$\tau=0.626$ (solid line), $\tau=1.126$ (dashed line),
$\tau=1.626$ (dotted line), and $\tau=2.126$ (dashed-dotted
line), where $\tau_1<\tau<\tau_2$. Other parameters are the
same as for Fig. 19. There exist points of singularity
$t_c^*$, defined by $a+bx(t_c^*-\tau)=0$, where
$x(t)|_{t\ra t_c^*-0}=+\infty$. These points are:
$t_c^*=11.4328$ (for solid line), $t_c^*=2.87170$ (for
dashed line), $t_c^*=3.33026$ (for dotted line), and
$t_c^*=3.83074$ (for dashed-dotted line).}
\label{fig:Fig.21}
\end{figure}

\newpage

\begin{figure}[h]
\centerline{\includegraphics[width=16cm]{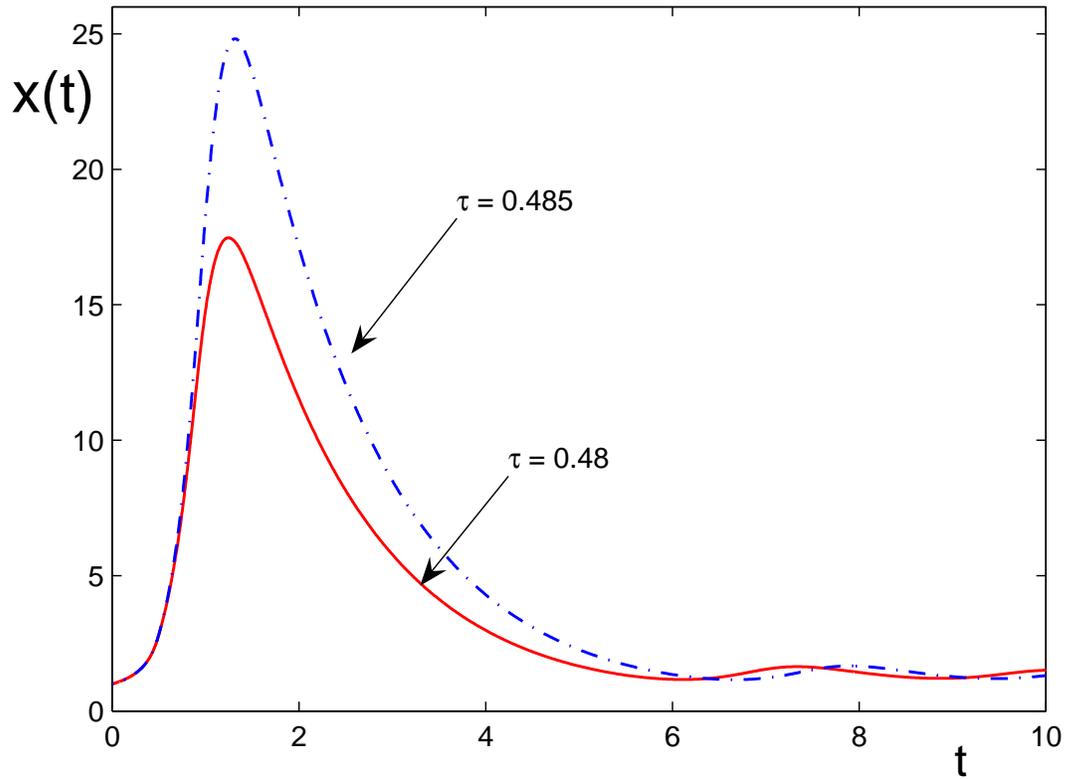}}
\caption{Solutions $x(t)$ to Eq. (\ref{eq61}) as
functions of time for the parameters $\tau=0.48$ (solid
line) and $\tau=0.485$ (dashed-dotted line), where
$\tau<\tau_0=0.495125$. Other parameters are: $a=2$,
$x_0=1$, and $b=-2.5$. The solutions $x(t)$ tend
non-monotonically to their stationary point
$x_2^*=-a/(b+1)=4/3$ as $t\ra\infty$.}
\label{fig:Fig.22}
\end{figure}

\newpage

\begin{figure}[h]
\centerline{\includegraphics[width=16cm]{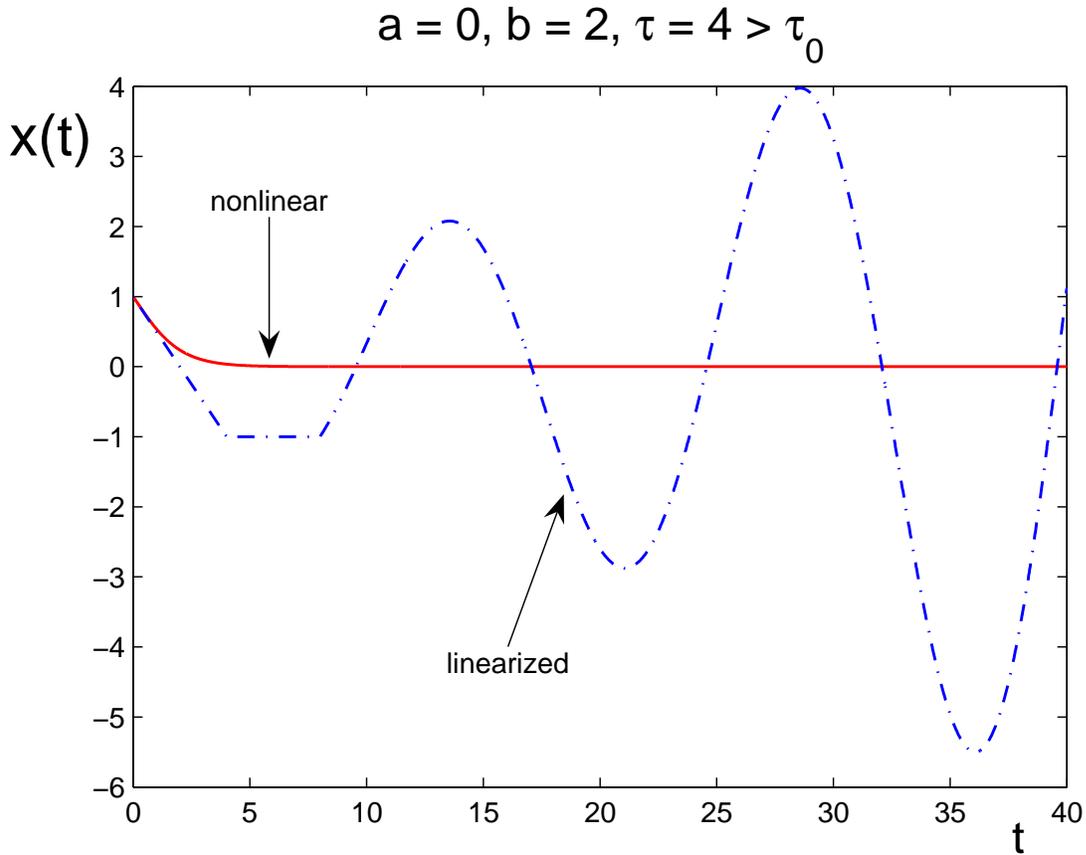}}
\caption{Solutions $x(t)$ to Eq. (\ref{eq81}) (solid
line) and to the corresponding equation obtained by linearizing
Eq. (\ref{eq81}) around the fixed
point $x^*=0$ (dashed-dotted line). Parameters for
the equations are: $a=0$, $b=2$, and $\tau=4$ (note that
$\tau>\tau_0=\pi$). The figure shows that the solution to the
linearized equation is unstable for $\tau>\tau_0$, as the
stability analysis prescribes, whereas the solution to the
nonlinear equation is stable for $\tau>\tau_0$, tending to
its stationary point $x^*=0$ as $t\ra\infty$.}
\label{fig:Fig.23}
\end{figure}

\newpage

\begin{figure}[h]
\centerline{\includegraphics[width=16cm]{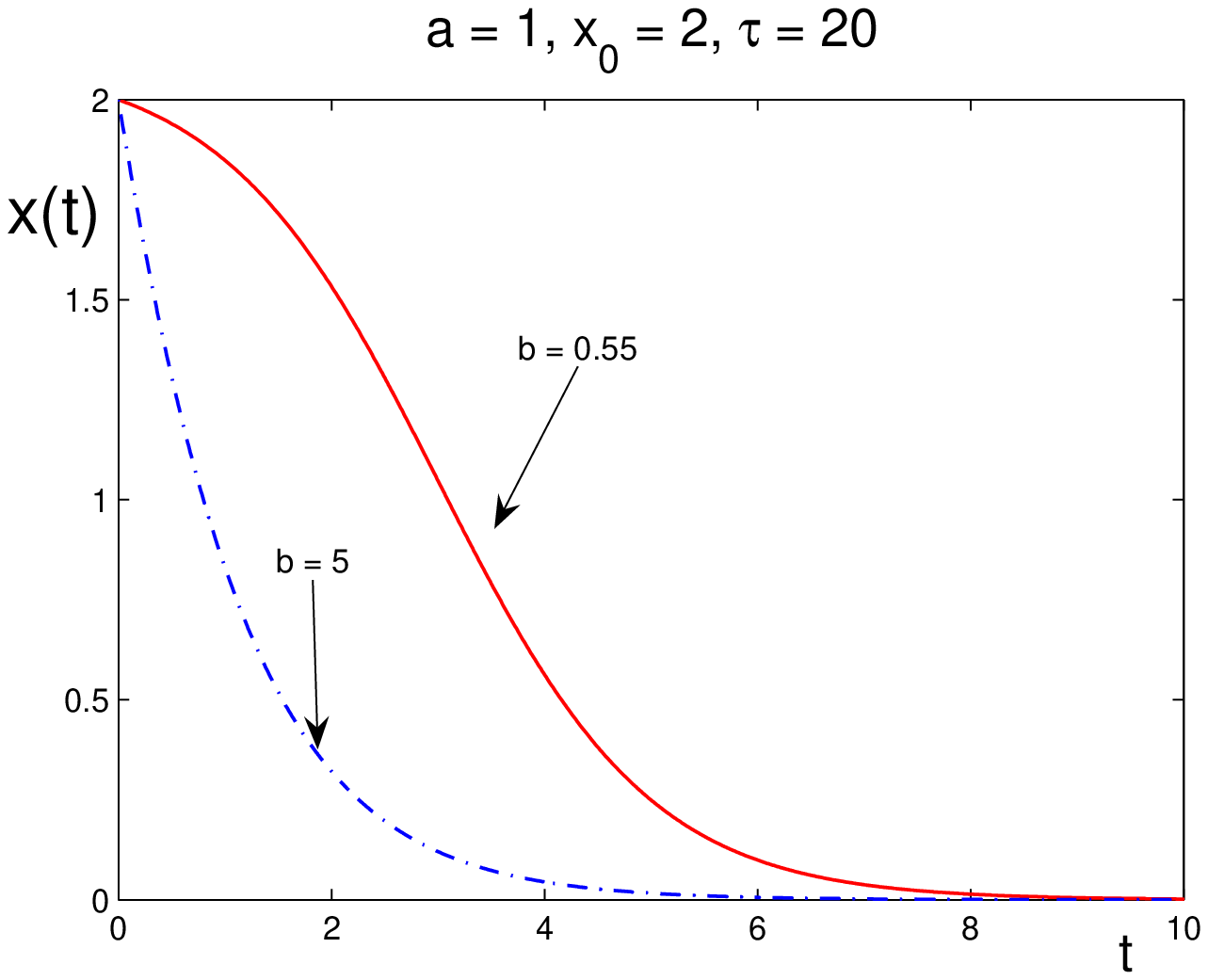}}
\caption{Solutions $x(t)$ to Eq. (\ref{eq81}) as
functions of time for the parameters $b=0.55$ (solid
line) and $b=5$ (dashed-dotted line). Other parameters
are: $a=1$, $x_0=2$, and $\tau=20$. The solutions $x(t)$
decrease monotonically to their stationary point $x^*=0$
as $t\ra\infty$.}
\label{fig:Fig.24}
\end{figure}

\newpage

\begin{figure}[h]
\centerline{\includegraphics[width=16cm]{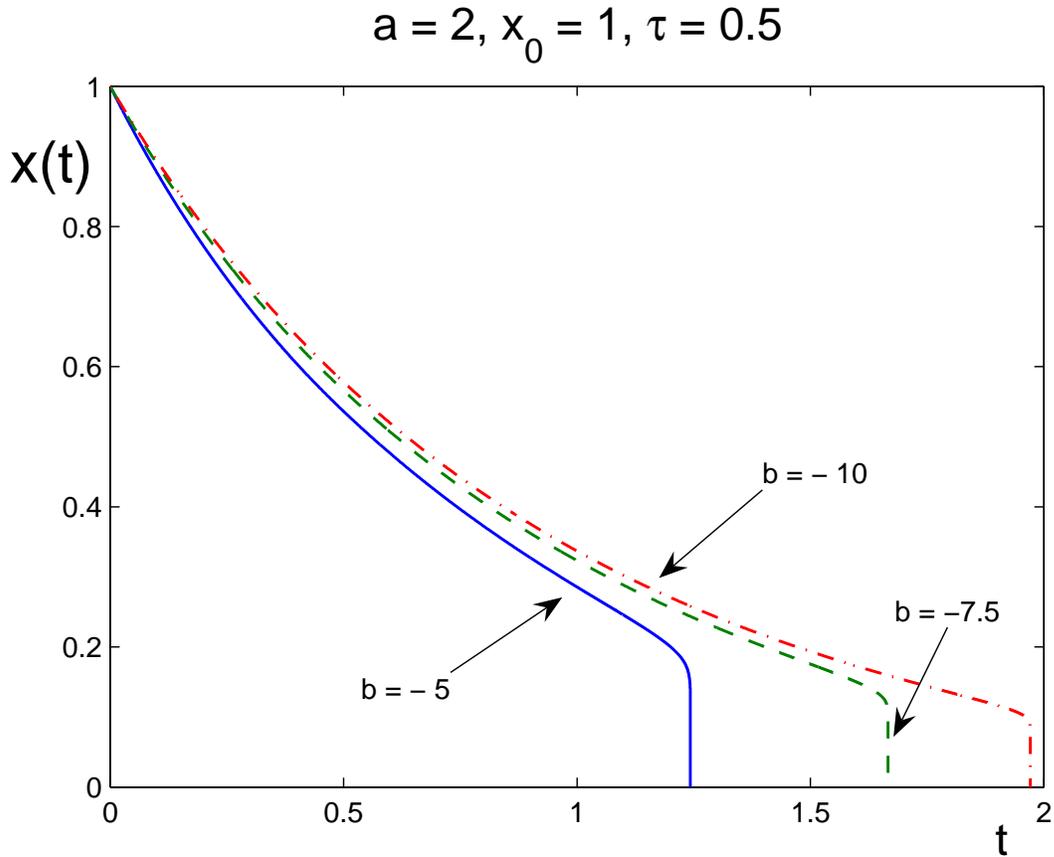}}
\caption{Solutions $x(t)$ to Eq. (\ref{eq81}) as
functions of time for the parameters $b=-5$ (solid line),
$b=-7.5$ (dashed line), and $b=-10$ (dashed-dotted line).
Other parameters are: $a=2$, $x_0=1$, and $\tau=0.5$.
The solutions $x(t)$ decrease monotonically with a sharp
but continuous drop ending at $0$ at time $t_d=1.24290$ (for
solid line), $t_d=1.66635$ (for dashed line), and $t_d=1.97114$
(for dashed-dotted line) defined by the equation
$a+bx(t_d-\tau)=0$. At these moments, of time
$\dot{x}(t)|_{t\ra t_d-0}=-\infty$.}
\label{fig:Fig.25}
\end{figure}

\newpage

\begin{figure}[h]
\centerline{\includegraphics[width=16.5cm]{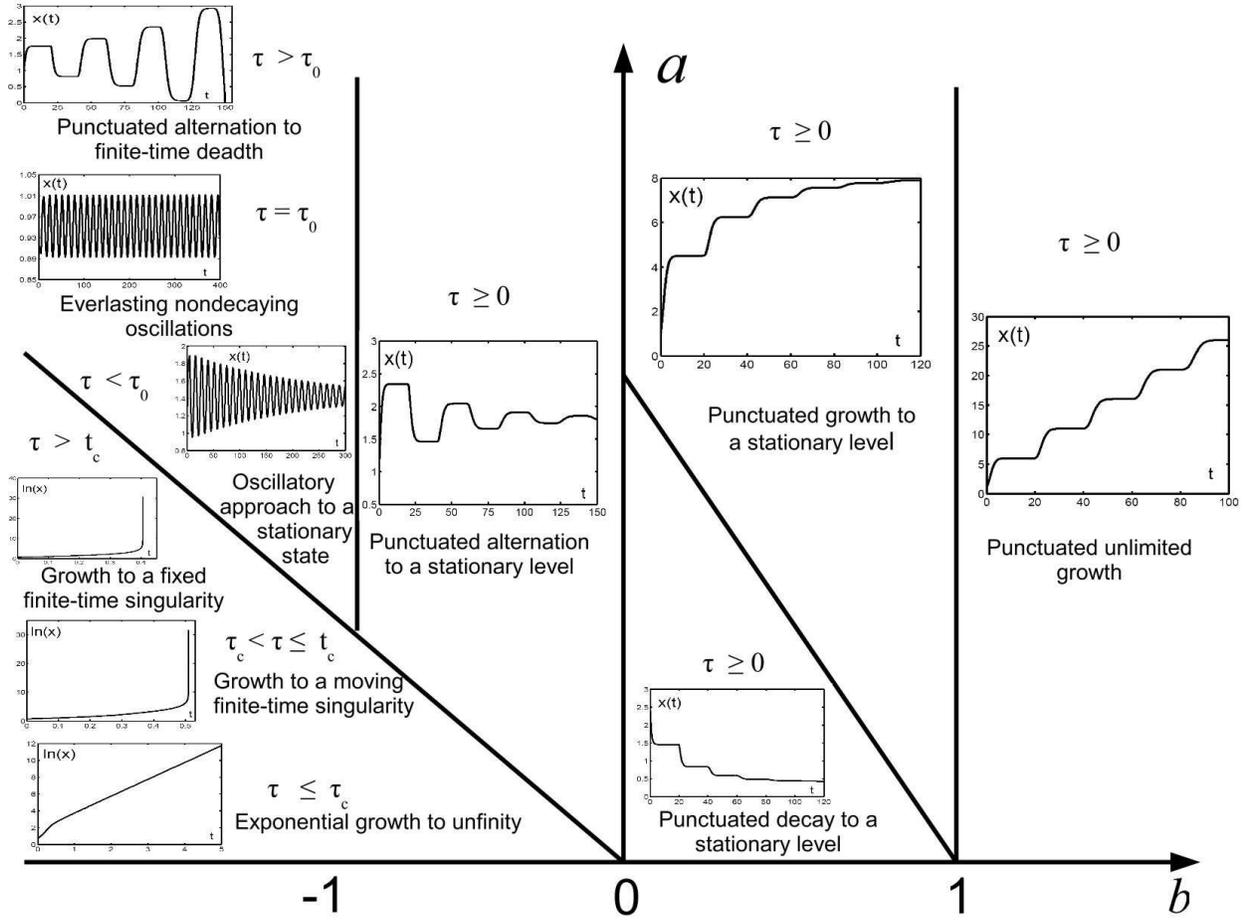}}
\caption{Scheme of the variety of qualitatively different 
solution types for the most complicated and the most realistic 
regime of Sec. 4, when gain (birth) prevails over loss (death) 
and competition is stronger than cooperation. } 
\label{fig:Fig.26}
\end{figure}

\newpage

\begin{figure}[h]
\centerline{\includegraphics[width=16.5cm]{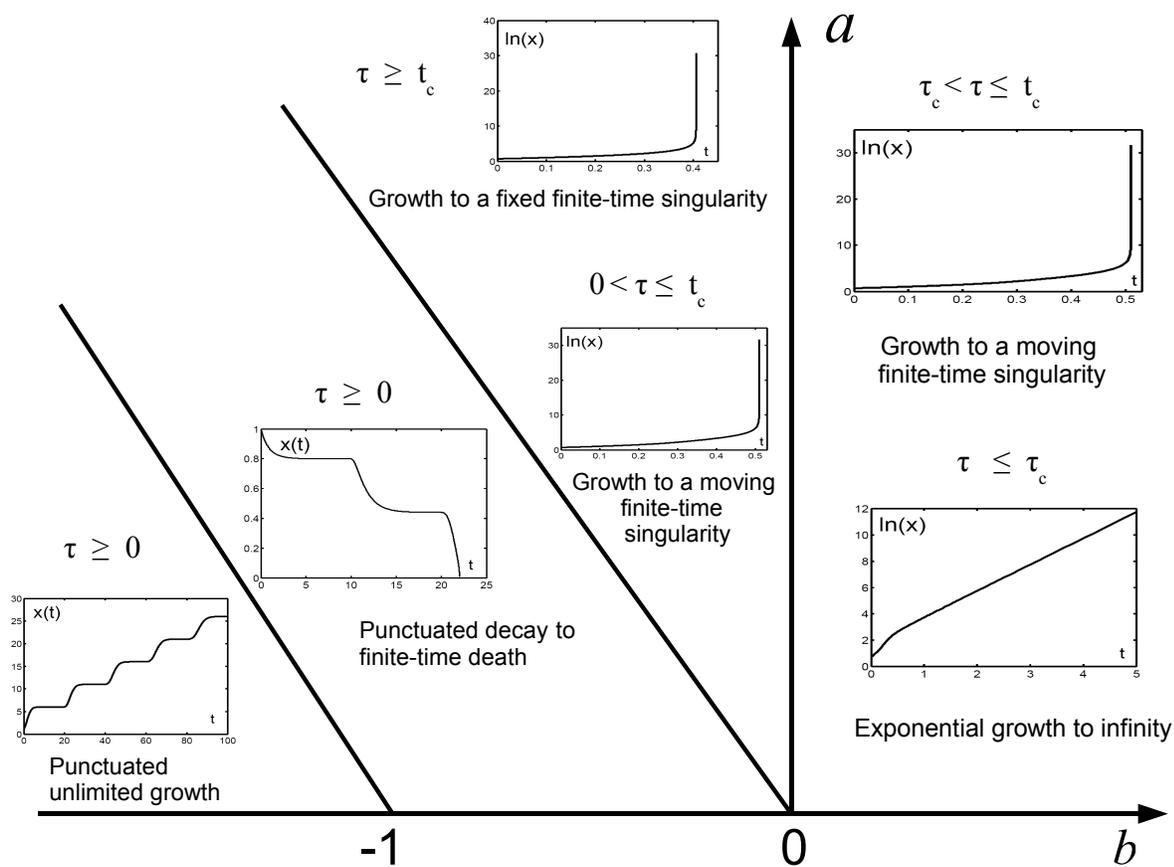}}
\caption{Scheme of qualitatively different solution types
for the case of Sec. 5, when gain prevails over loss and 
cooperation prevails over competition.} 
\label{fig:Fig.27}
\end{figure}

\newpage

\begin{figure}[h]
\centerline{\includegraphics[width=16.5cm]{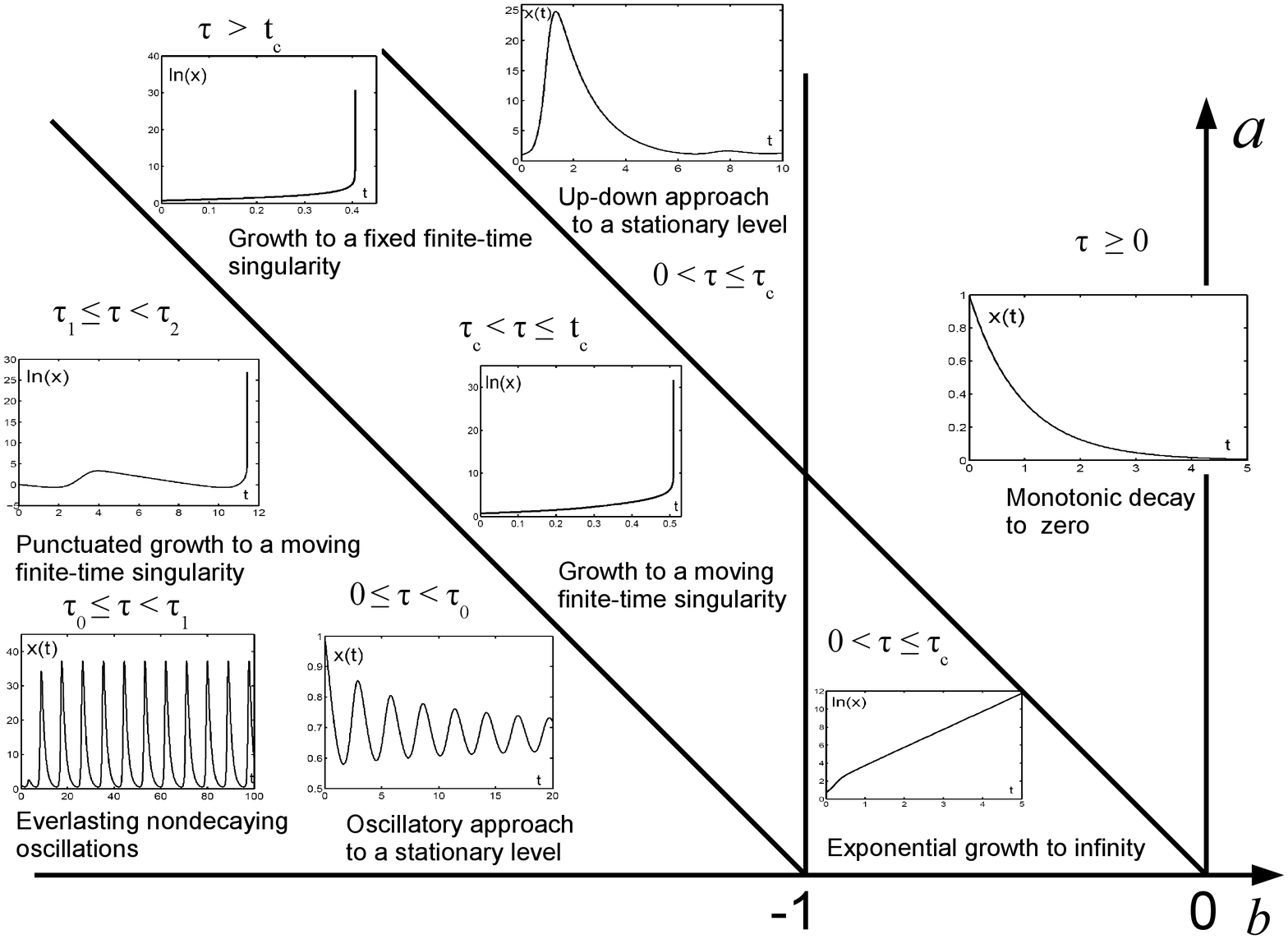}}
\caption{Summarizing scheme of qualitatively different solution 
types for the case of Sec. 6, when loss prevails over gain and 
competition prevails over cooperation.} 
\label{fig:Fig.28}
\end{figure}

\newpage

\begin{figure}[h]
\centerline{\includegraphics[width=16.5cm]{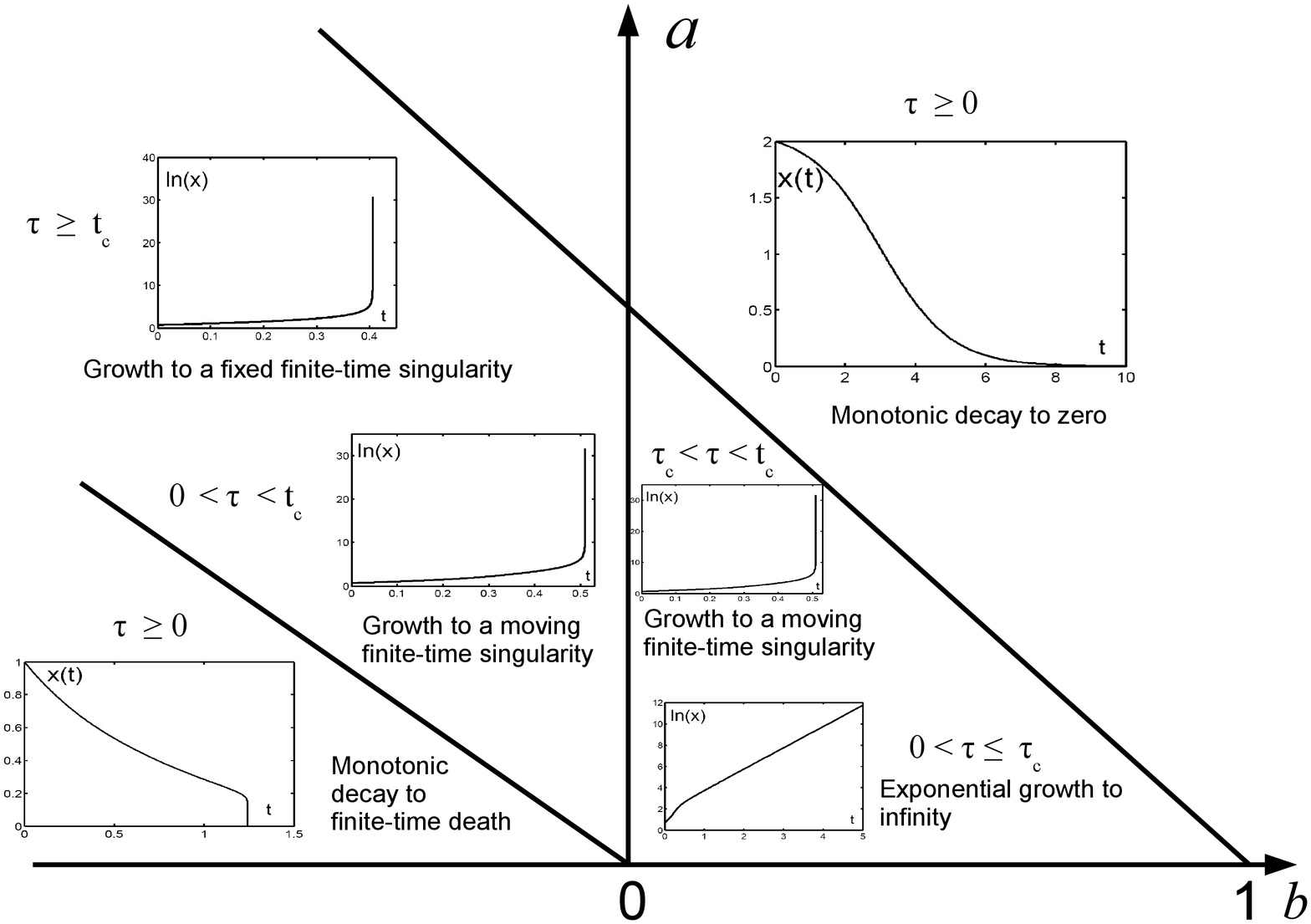}}
\caption{Summarizing scheme of qualitatively different solution 
types for the case of Sec. 7, when loss prevails over gain and 
cooperation prevails over competition.} 
\label{fig:Fig.29}
\end{figure}


\newpage

\begin{thebibliography}{99}

\bibitem{13}
A. Korotayev, in: History and Mathematics: Historical
Dynamics and Development of Complex Societies, P. Turchin,
ed., Komkniga, Moscow, 2007, p. 44-62.

\bibitem{14}
L.M. Bettencourt, J. Lobo, D. Helbing, C. Kuhnert, G.B. West, Proc.
Natl. Acad. Sci. 104 (2007) 7301-7306.

\bibitem{15}
T. Modis, Futurist N5 (2003) 26-31.

\bibitem{16}
T. Modis, Technol. Forecast. Soc. Change 47 (1994) 63-73.

\bibitem{17}
N. Eldredge, S.J. Gould, in: Models in Paleobiology, T.J. Schopf,
ed., Freeman, San Francisco, 1972, p. 82-115.

\bibitem{18}
S.J. Gould, Natural Hist. 86 (1977) 12-16.

\bibitem{19}
S.J. Gould, Science 216 (1982) 380-387.

\bibitem{20}
N. Eldredge, Time Frames, Simon and Schuster, New York, 1985.

\bibitem{21}
S.J. Gould, The Structure of Evolutionary Theory, Harvard
University, Cambridge, 2002.

\bibitem{22}
E. Meyer, Animal Species and Evolution, Harvard University,
    Cambridge, 1963.

\bibitem{23}
E. Meyer, Evolution and the Diversity of Life, Belknap, Cambrodge,
    1976.

\bibitem{24}
E. Meyer, The Growth of Biological Thought, Harvard University,
    Cambridge, 1982.

\bibitem{25}
E. Meyer, One Long Argument, Harvard University, Cambridge, 1991.

\bibitem{26}
D.R. Prothero, Skeptic 1 (1992) 38-47.

\bibitem{27}
C. Darwin, On the Origin of Species, Murray, London, 1859.

\bibitem{28}
H. Simon, Quarter. J. Econ. 69 (1955) 99-118.

\bibitem{29}
M. Givel, Policy Stud. J. 43 (2006) 405-418.

\bibitem{30}
C.J. Gersick, Acad. Manag. J. 31 (1988) 9-41

\bibitem{31}
C.J. Gersick, Acad. Manag. Rev. 16 (1991) 10-36.

\bibitem{32}
H. Arrow, M.S. Poole, K.B. Henry, S. Wheelan, R. Moreland,
    Small Group Res. 35 (2004) 73-105.

\bibitem{33}
J.H. Wilmore, D.L. Costill, W.L. Kenney, Physiology of Sport
and Exercise, Human Kinetics, Champaign, 2008.


\bibitem{1}
P.F. Verhulst, Corr. Math. Phys. 10 (1838) 113-117.

\bibitem{2}
 G.E. Hutchinson, Ann. NY Acad. Sci. 50 (1948) 221-246.

\bibitem{3}
 L. Jingwen, Appl. Math. B 11 (1996) 165-174.

\bibitem{4}
 C. Marchetti, P.S. Meyer, J.H. Ausubel, Technol. Forecast. Soc.
   Changes 52 (1996) 1-30.

\bibitem{5}
Y. Louzoun, S. Solomon, J. Goldenberg, D. Mazursky, Artif.
Life 9 (2003) 357-370.

\bibitem{6}
 J.P. Gabriel, F. Saucy, L.F. Bersier, Ecol. Model. 185
(2005) 147-151.

\bibitem{7}
J. Hui, L. Chen, IMA J. Appl. Math. 70 (2005) 479-487.

\bibitem{8}
M. Berezowski, E. Fudala, Chaos Solit. Fract. 28 (2006) 543-554.

\bibitem{9}
J. Arino, L. Wang, G.S. Wolkowicz, J. Theor. Biol. 241 (2006)
109-119.

\bibitem{10}
K. Gopalsamy, Stability and Oscillations in Delay Differential
   Equations of Population Dynamics, Kluwer, Dordrecht, 1992.

\bibitem{11}
V. Kolmanovskii, A. Myshkis, Introduction to the Theory and
Applications of Functional Differential Equations, Kluwer,
Dordrecht, 1999.

\bibitem{12}
O. Arino, M.L. Hbid, E.A. Dads, eds., Delay Differential
Equations and Applications, Springer, Dordrecht, 2006.


\bibitem{35}
W.F. Laurence, Biol. Conserv. 91 (1999) 109-117.

\bibitem{36}
S.L. Pimm, P. Raven, Nature 403 (2000) 843-845.

\bibitem{37}
E.W. Sanderson, M. Jaiteh, M.A. Levy, K.H. Redford, A.V. Wannebo,
G. Woolmer, Biosciences 52 (2002) 891-904.

\bibitem{38}
J.K. McKee, P.W. Sciulli, C.D. Fooce, T.A. Waite, Biol. Conserv.
    115 (2003) 161-164.

\bibitem{39}
W.M. Hern, Populat. Environm. 21 (1999) 59-80.

\bibitem{40}
D. Sornette, Phys. Rep. 378 (2003) 1-98.

\bibitem{41}
D. Sornette, Why Stock Markets Crash, Princeton University,
Princeton, 2003.

\bibitem{34}
D. Darcet, D. Sornette, J. Econ. Interact. Coord. 3 (2008)
137-163.

\bibitem{42}
S.D. Brierly, J.N. Chiasson, E.B. Lee, S.H. Zak, IEEE Trans.
Automat. Control. 27 (1982) 252-254.

\bibitem{43}
J.K. Hale, E.F. Infante, F.P. Tsen, J. Math. Anal. Appl. 105
(1985) 533-555.

\bibitem{44}
V.B. Kolmanovskii, L. Torelli, V. Vermiglio, SIAM J. Math.
Anal. 25 (1994) 948-961.

\bibitem{45}
J. Chen, IEEE Trans. Automat. Control. 40 (1995) 1087-1093.

\bibitem{46}
X. Mao, Syst. Contr. Lett. 28 (1996) 159-165.

\bibitem{47}
J. Losson, M.C. Mackey, A. Longtin, Chaos 3 (1993) 167-176.

\bibitem{48}
J.D. Farmer, Physica D 4 (1982) 366-393.

\bibitem{49}
R. Cowen, The History of Life, Blackwell Science, Cambridge,
1994.

\bibitem{50}
R. Leakey, R. Lewin, La Sixi\'{e}me Exctinction, Flammarion,
Paris, 1999.

\bibitem{51}
R.J. Sawyer, Calculating God, Tom Doherty, New York, 2000.

\bibitem{52}
E.O. Wilson, The Future of Life, Random House, New York, 2002.

\bibitem{53}
R.A. Rhonde, R.A. Muller, Nature 434 (2005) 209-210.

\bibitem{54}
P.D. Ward, Under Green Sky, Harper Collins, New York, 2007.

\bibitem{55}
V.I. Yukalov, A.S. Shumovsky, Lectures on Phase Transitions,
World Scientific, Singapore, 1990.

\bibitem{56}
V.I. Yukalov, Phys. Rep. 208 (1991) 395-492.

\bibitem{57}
D. Sornette, Critical Phenomena in Natural Sciences, Springer,
Berlin, 2004.

\bibitem{58}
A. Johansen, D. Sornette, Physica A 294 (2001) 465-502.

\bibitem{59}
J.H. Ausubel, P.S. Meyer, Human Dimens. Quart. 1 (1994) 17-19.

\bibitem{60}
BP Statistical Review of World Energy, London, 2008.

\bibitem{61}
D.K. Ginther, S. Kahn, J. Econ. Perspect. 18 (2004) 193-214.

\bibitem{62}
A.Y. Yakovlev, K. Boucher, M. Mayer-Proschel, M. Noble, 
Proc. Natl. Acad. Sci. USA 95 (1998) 14164-14167.

\bibitem{63} 
J. Balzarini, B. Degreve, S. Hatse, E. De Clerco, M. Breuer, 
M. Johansson, R. Huybrechts, A. Karlsson, Molec. Pharmacol. 57 
(2000) 811-819.

\bibitem{64}
N. Cougot, S. Babajko, B. Seraphin, J. Cell Biol. 165 (2004) 31-40.

\bibitem{65}
C. Marchetti, P.S. Meyer, J.H. Ausubel, Technol. Forecast. Social 
Change 52 (1996) 1-30.


\end{thebibliography}
\end{document}